\documentclass[amssymb,amsmath,aps,12pt,tightenlines,superscriptaddress,longbibliography,notitlepage]{revtex4-1}

\usepackage{graphicx,mdwlist,color,wrapfig,revsymb}
\usepackage{etoolbox}

\begin{document}

\title{New topological excitations and melting transitions in quantum Hall systems}

\author{Tzu-ging Lin}
\affiliation{Department of Physics, Purdue University, West Lafayette, Indiana 47907 USA}
\author{George Simion}
\affiliation{Department of Physics, Purdue University, West Lafayette, Indiana 47907 USA}
\author{John D. Watson}
\affiliation{Department of Physics, Purdue University, West Lafayette, Indiana 47907 USA}
\affiliation{Birck Nanotechnology Center, Purdue University, West Lafayette, Indiana 47907 USA}
\author{Michael J. Manfra}
\affiliation{Department of Physics, Purdue University, West Lafayette, Indiana 47907 USA}
\affiliation{Birck Nanotechnology Center, Purdue University, West Lafayette, Indiana 47907 USA}
\affiliation{Department of Electrical and Computer Engineering, Purdue University, West Lafayette, Indiana 47907 USA}
\affiliation{Department of Material Engineering, Purdue University,
West Lafayette, Indiana 47907 USA}
\author{Gabor A. Csathy}
\affiliation{Department of Physics, Purdue University, West Lafayette, Indiana
47907 USA}
\author{Yuli Lyanda-Geller}
\thanks{To whom correspondence on theory should be addressed. E-mail: yuli@purdue.edu}
\affiliation{Department of Physics, Purdue University, West Lafayette, Indiana
47907 USA} \affiliation{Birck Nanotechnology Center, Purdue University, West
Lafayette, Indiana 47907 USA}
\author{Leonid P. Rokhinson}
\thanks{To whom correspondence on experiment should be addressed. E-mail: leonid@purdue.edu}
\affiliation{Department of Physics, Purdue University, West Lafayette, Indiana
47907 USA}\affiliation{Birck Nanotechnology Center, Purdue University, West
Lafayette, Indiana 47907 USA} \affiliation{Department of Electrical and Computer Engineering, Purdue University,
West Lafayette, Indiana 47907 USA}


\begin{abstract}
We discover a new topological excitation of two dimensional electrons in
 the quantum Hall regime. The strain dependence of resistivity is shown to
 change sign upon crossing filling-factor-specified boundaries of reentrant
 integer quantum Hall effect (RIQHE) states. This observation violates the
 known symmetry of electron bubbles thought to be responsible for the
 RIQHE. We demonstrate theoretically that electron bubbles become elongated
 in the vicinity of charge defects and form textures of finite size.
 Calculations confirm that texturing lowers the energy of excitations.
 These textures form hedgehogs (vortices) around defects having (lacking)
 one extra electron, resulting in striking strain-dependent resistivity
 that changes sign on opposite boundaries of the RIQHE. At low density
 these textures form an insulating Abrikosov lattice.  At densities
 sufficient to cause the textures to overlap, their interactions are
 described by the XY-model and the lattice melts.  This melting explains
 the sharp metal-insulator transition observed in finite temperature
 conductivity measurements.
\end{abstract}

\maketitle

Topology and symmetry define states of matter and their responses to external
forces. How solids melt and become fluids, or how insulators become
conductors is often controlled by the generation of topological excitations.
In a magnetic field the energy spectrum of electrons is quantized into
discrete Landau levels, causing a two-dimensional electron gas transitions
through a variety of ground states as a function of magnetic field. When an
integer number of single-particle energy levels are filled (filling factor
$\nu=\Phi_0/\Phi$ is integer, where $\Phi_0=h/e$ is a flux quanta, $\Phi=B/n$
is magnetic flux per electron, $B$ is magnetic field and $n$ is electron
density), the Hall resistance becomes quantized in multiples of $R_q=h/e^2$
and longitudinal resistance vanishes, the hallmarks of the integer quantum
Hall effect\cite{Klitzing1980}. Electron-electron interactions dominate
physics at low filling factors ($\nu<1$) and the many-body ground state is a
competition between fractional quantum Hall effects\cite{tsui82}, and Wigner
crystals\cite{Wigner1934}. In the 2nd Landau level ($2<\nu<4$) exotic
even-denominator fractional quantum Hall states, possibly possessing
non-Abelian excitations, are found \cite{moore91}. On the 3rd and higher
Landau levels ($\nu>4$) a unidirectional charge density wave, the stripe
phase, is formed near half-integer filling
factors\cite{koulakov96,du99,lilly99}. Between integer and half-integer
filling factors another set of states characterized by Hall resistance
quantized to the nearest integer value but occurring at non-integer filling
factor is observed, the so-called reentrant integer quantum Hall effect
(RIQHE) states\cite{Eisenstein2001,Eisenstein2001a}. Here we report discovery
of a new class of charged topological excitations, realized in two
dimensional electron systems in the RIQHE, in which elongated bubbles of
electrons form finite-size textures around charge defects.

In a two-electron (2\={e}) bubble crystal there are one-
and three-electron (1\={e} \& 3\={e}) charge defects. We propose that elongated
bubbles form hedgehogs-like and vortex-like textures around 1\={e} \& 3\={e}
respectively. We show that formation of textures considerably reduce energy of charged
defects. Remarkably, these topological defects are
very sensitive to uniaxial strain and strain-dependent resistivity has
different sign for vortex and hedgehog textures. This unique property enables
detailed investigation of topological defects and a Berezinsky-Kosterlitz-Thouless-like (BKT) transition \cite{Berezinskii1972,Kosterlitz1973} in the
previously unexplored regime, where the relative density of topological defects
with different symmetry can be tuned by magnetic field. In this regime
melting is a function of several variables and forms a continuous phase
boundary in the field-temperature ($B-T$) plane.
At low densities these textures do not overlap. They are bound in an insulating crystal \cite{Abrikosov,Wigner1934}.
At high defect density, driven either by finite temperature or magnetic field, the textures begin to overlap and their interactions
define the energetics of the system.
These interactions are described by XY-model and induce melting of the defect lattice. The melting described here differs from the standard
BKT transition \cite{Berezinskii1972,Kosterlitz1973} because these finite-size topological textures must have significant spatial overlap to induce vortex unbinding.
This melting transition resembles asymptotic freedom of quarks requiring
them to be "squeezed" in order to be freed\cite{Gross1973}. Melting of the
lattice of topological defects explains the sharp boundary between an insulating
RIQHE state and a conducting
state observed in a recent experiment\cite{Deng2012b}.

Magnetoresistance in a high mobility 2D electron gas is
shown in Fig.~\ref{f-anis} for filling factors between $\nu=4$ and 5. A well-defined stripe phase with large resistance anisotropy between $[110]$ and $[1\bar{1}0]$ crystallographic directions is seen near $\nu=4.5$. RIQHE phases are formed near $\nu=4.3$ and 4.7 and are marked as $B_1$ and $B_2$ on the plot. The width of RIQHE phases is temperature dependent,  see Fig.~\ref{f-T}, with the insulating state bounded by a sharp
resistance peak which separates insulating and conducting phases in the
($B$-$T$) plane. The most striking feature shown in Fig.~\ref{f-anis} is strain dependence of resistivity. Uniaxial strain is expected to introduce slight anisotropy of the exchange interaction and of the effective mass for electrons in GaAs. Indeed, for the stripe phase we see small enhancement of resistivity in $[1\bar{1}0]$ and reduction in $[1\bar{1}0]$ consistent with expectations\cite{Koduvayur2011}. Unexpectedly, however, the resistance vs. strain dependence changes sign across the RIQHE states: the sign of the change of resistance with strain near $\nu=4.33$ and 4.66 is consistent with the induced anisotropy of the exchange, while near $\nu=4.23$ and 4.73 the sign of change is reversed, see Fig.~\ref{f-T}.

The RIQHE state was originally suggested to be a crystal of uniformly charged
bubbles holding integer number of
electrons\cite{fogler96,Fogler1997,Goerbig2003,Cote2003,Ettouhami2006,fogler-review}.
Electron hopping between bubbles is forbidden by Coulomb blockade and
conductance vanishes as long as the bubble crystal is pinned by disorder. It is conceivable that the
high temperature transition to a conducting state takes place due to trivial melting
of the bubble crystal. However, the observed strain response of resistivity is incompatible with the symmetry of the bubble crystal, a bubble
liquid or a uniform electron liquid, and thus, the observed metal-insulator transition cannot be atributed to the melting of the bubble crystal. In the presence of strain $\varepsilon$
resistivity becomes anisotropic, $\varrho_{ii}=\varrho^{0}_{ii} +
\beta(\varepsilon_{ii}-\varepsilon_{jj})]$,  and the sign of the constant $\beta$
is determined by the symmetry of the underlying state. This
intrinsic parameter $\beta$ has to change sign across the RIQHE state as a
function of magnetic field to be consistent with the experimentally measured strain dependence of resistivity. This requires a symmetry change across the boundary of the
RIQHE state.

In order to gain insight into the physics of RIQHE, it is instructive to
analyze the effect of strain on resistivity from the vantage point afforded by
symmetry invariants. Finite conductivity is due to movement of charge defects, where an
extra electron on a bubble (or a lack of one) can hop through the bubble
lattice. We write down the expression for electric current in the ohmic
approximation using a real space approach, $\mathbf{j}=\int \mathbf{r}
(e\mathbf{E}\mathbf{r})f(\mathbf{r})dr$, where $\mathbf{E}$ is the electric
field, $\mathbf{r}$ is the electron coordinate operator, and $f(\mathbf{r})$
describes the distribution of charge and reflects the presence of strain and other
symmetry factors. For this symmetry analysis it is sufficient to consider $\mathbf{r}$ continuous;  the results are the same as for consideration of
hopping over discrete lattice sites. Resistivity $\varrho_{ii}$ is defined by
the spatial average $\left<r_i^2f(\mathbf{r})\right>$.   The function
$f(\mathbf{r})$  can be conveniently written as a product of symmetry
invariants. The symmetry invariant
$T_1=(\varepsilon_{xx}-\varepsilon_{yy})(\mathbf{r}_x^2-\mathbf{r}_y^2)$
describes electron hopping in an ideal lattice deformed by strain and motion of
non-textured charge defects, as shown in Figs.~\ref{fsym_1}(a,b). Anisotropy of
$f(\mathbf{r})\propto T_1$ is similar to the strain-induced anisotropy of the
effective mass: the sign of corresponding $\beta$ would be the same across all
RIQHE regions in contrast to the experimental data. However, we discover that in
the vicinity of charge defects bubbles change their shape from round to
elongated, which results in the formation of non-trivial textures,
Fig.~\ref{fsym_1}c. We identify two possible textures: vortices and 2D
hedgehogs described by functions $T_2^{(v)}= (\mathbf{a}\times
\mathbf{r})_z^2$ and $T_2^{(h)}= (\mathbf{a}\cdot \mathbf{r})^2$, where vector
$\mathbf{a}$ is the elongation of a bubble. In the presence of bubble
elongations, strain affects the system via the symmetry invariant
$T_3=(\varepsilon_{xx}-\varepsilon_{yy})(\mathbf{a}_x^2-\mathbf{a}_y^2)$.
Calculating the resistivity $\left<r_i^2f(\mathbf{r})\right>$ by averaging over
$\mathbf{a}$ for $f^{(h)}(\mathbf{r})\propto T_2^{(h)}T_3$ and
$f^{(v)}(\mathbf{r})\propto T_2^{(v)}T_3$, we arrive to the opposite sign of
$\beta$ for vortices and hedgehogs. Below we show that a microscopic calculation
of resistivity shows the sign change as seen in experiment. Vortex topological excitations dominate
on one side of RIQHE, while hedgehogs dominate on the other side, which results
in the symmetry change across the RIQHE necessary for the change of the sign of
$\beta$. We note here that from the vantage point afforded by symmetry
$\varepsilon_{xx}-\varepsilon_{yy}$ behaves exactly like an expression $H_x^2-H_y^2$
in the presence of tilted magnetic field, where $H_x$ and $H_y$ are the in-plane components of the field.
Therefore it is possible that experiments in tilted field will also reveal opposite effects on resistivity  due to
vortices and hedgehogs.

To describe microscopic properties of the the system we use the real
space approach of ``interacting guiding centers" proposed by
Ettouhami\cite{Ettouhami2006}. Throughout the rest of the paper distances are
expressed in units of magnetic length $\lambda$, wavevectors in units of
$1/\lambda$ and energies in units of $e^2/\kappa\lambda$, where $\kappa$ is the
background dielectric constant. In this approach, the Hartree-Fock (HF)
Hamiltonian is the sum of effective interactions between guiding centers:
$H_{\rm{HF}}=1/2( \sum_{i\neq j}U({\bf R}_{ij})+\sum_i U(0)$) where $i$ and $j$
labels the nodes of triangular lattice, $\mathbf{R}_i$ is the coordinate of the
bubble $i$, ${\bf R}_{ij}={\bf R}_{i}-{\bf R}_{j}$. An effective interaction
between bubbles $i$ and $j$ is given by
\begin{equation}
\label{eq:Ueff_gen} U({\bf{R}}_{ij})=\int{\frac{d^2\bf{q}}{(2\pi)^2}
\rho_{i}^*({\bf {q}}) [V_{\rm{H}}({\bf{q}})-V_{\rm{F}}(\bf{q})]}
\rho_{j}({\bf {q}})\exp{(i\mathbf{q}\cdot{\bf R}_{ij})}~,
\end{equation}
where $\rho_{i}$ represents the density of the bubble located at site $i$
projected on the uppermost Landau Level. The Hartree potential and exchange
potentials are, respectively \cite{aleiner95}, $V_{\rm{H}}({\bf{q}})=\frac{2
\pi}{q}e^{-q^2/2}[L_{n}(q^2/2)]^2$, and
$V_{\rm{F}}({\bf{q}})=2\pi\int{\frac{d^2\bf{q'}}{(2\pi)^2}V_{{\rm{H}}}({\bf{q}})
e^{-i ({\bf{q}}\times {\bf{q'}})\cdot\hat z}}$, where $L_{n}$ is the
$n^{\rm{th}}$ Laguerre polynomial. To describe the experimental data we
concentrate on a two-electron bubble crystal and its excitations. The ground
state of an ideal bubble crystal at zero temperature is characterized by
round bubbles. Calculated energies of isolated 1\={e} and 3\={e}
defect in a crystal of round bubbles taking into account displacements of lattice bubbles due to defect
are plotted in Fig.~\ref{f-defects}
with dashed lines. Apart from the symmetry of these defects being inconsistent
with the strain dependence of resistivity, their activation energies are in the
range of several Kelvin and, thus, cannot explain high conductivity measured around
100 mK.

The presence of charged defects induces elongation of bubbles in the vicinity of
defects. The wavefunction of an elongated two-electron bubble with guiding
centers separated by $2\mathbf{a}$ is:
\begin{equation}
\label{eq:wf_2e_bubble_gen}
\Psi_{\xi,\bf{a}}({\bf{r}}_1,{\bf{r}}_2)=\alpha \left(\psi_{\bf{a}}({\bf{r}}_1,{\bf{r}}_2)
+\xi\psi_{-\bf{a}}({\bf{r}}_1,{\bf{r}}_2)\right)
\end{equation}
where $\alpha$ is the normalization coefficient,
\[\psi_{\bf{a}}({\bf{r}}_1,{\bf{r}}_2)=[u_{-2}({\bf{r}}_1+{\bf{a}})
u_{-1}({\bf{r}}_2-{\bf{a}})e^{i({\bf{r}}_1-{\bf{r}}_2)\times
{\bf{a}}/2}-u_{-1}({\bf{r}}_1-{\bf{a}})
u_{-2}({\bf{r}}_2+{\bf{a}})e^{i({\bf{r}}_2-{\bf{r}}_1)\times{\bf{a}}/2}], \]
$u_{-2}$ and $u_{-1}$ are single-electron wavefunctions on the 3rd Landau
level (3LL) in the symmetric gauge with angular momentum projections $m=-2$ and
$m=-1$ \cite{Giuliani-book}. This is a trial variational wavefunction, similar in
spirit to the one proposed by Fogler and Koulakov for round
bubbles\cite{Fogler1997}. The wavefunction
(\ref{eq:wf_2e_bubble_gen}) reflects both the magnitude and orientation of
bubble elongation.

Expressions for the electron density of an elongated bubble projected on the
3LL, $\rho_{\xi,\bf{a}}({\bf{q}})$, and interaction energy between of elongated
bubbles, $U_2({\bf{R}}_{i,j})$, are presented in the Supplemental Information.
Besides $U_2({\bf{R}}_{i,j})$, the total energy includes contributions from
charged defects. The wavefunction of a 1\={e} bubble is  $u_{-2}({\bf r})$ and
has round shape. The wavefunction of a 3\={e} bubble is a Slater determinant
made of $u_{-2}$, $u_{-1}$, and $u_{m=0}$. Exact structure and shape of these
defects can be determined by minimizing the cohesive energy, however, energetics
and electron transport are primarily determined by elongations of the
crystalline two-electron bubbles. Thus, the detailed structure of the 3\={e}
bubble is not essential and we model it as having all three guiding centers on
top of each other.  We now
calculate interactions of a single charge defect at a site $k$
with 2\={e} bubbles. It is convenient to express these interactions in terms of
elongations $\mathbf{a_i}$ and effective dipole moments
$\mathbf{\boldsymbol{\mu}_i(a_i)}$, which represent
rotated and rescaled $\mathbf{a_i}$
. Retaining terms up to $R_{ik}^{-3}$ we arrive to elongation-dependent part of
asymptotic expansion of energy:
\begin{eqnarray}
\label{eq:effH_pm_d}
 H^{\pm}&=&\sum_{i\neq
k}
\left[
\mp\left[\frac{\boldsymbol\mu_i\cdot \hat {\mathbf{R}}_{ik}}{R_{ik}^2}+
\frac{v_i}{2 R_{ik}^3} \cos{(2\varphi_i)}\right]+ U_f^{\pm}(i)\right]
\nonumber\\
&+&\frac{1}{2}\sum_{i \neq j}\frac{ {\boldsymbol{\mu}}_i \cdot {\boldsymbol{\mu}}_j- 3
({\boldsymbol{\mu}}_i \cdot \hat{{\mathbf{R}}}_{ij})({\boldsymbol{\mu}}_j \cdot
\hat{{\mathbf{R}}}_{ij}) } {R_{ij}^3}~,
\end{eqnarray}
where $(+)$ and $(-)$ signs correspond to 1\={e} and 3\={e} defects,
$\hat{{\mathbf{R}}}_{ij}=\mathbf{R_{ij}}/R_{ij}$ and $\varphi_i$ is the angle
between $\mathbf{a}_i$ and $\mathbf{R}_i$. The first three terms of Eq.
(\ref{eq:effH_pm_d}) are contributions to energy at the site $i$ due to the
presence of the defect, and the fourth term is an elongation-dependent
interaction between two sites $i$ and $j$. Explicit expressions for
$\mathbf{\mu_i}$, $v_i$,  and  $U_f^{\pm}(i)$,  are given in Supplementary
Information.

The ground state of the bubble system in the presence of elongations caused by a defect is obtained
by minimizing the energy (\ref{eq:effH_pm_d}). For 1\={e} defects the energy
minimum corresponds to $\varphi_i=0 ,\pi$, which we call vortices and
anti-vortices correspondingly, and for 3\={e} defect the energy minimum occurs
at $\varphi_i=\pi/2 ,3\pi/2$, which we will call hedgehogs and anti-hedgehogs
(although these are 2D objects). The two chiralities of vortex and hedgehog
textures are almost degenerate and the difference can be ignored for
experimentally relevant temperatures. The
activation energies of the defects in the presence of elongations of lattice bubbles are shown by
solid lines in  Fig.~\ref{f-defects}b. The resulting energies are 10 times
smaller than excitation energies for charge defects in the absence of textures,
and are consistent with experimentally studied range of temperatures. Density profiles of vortex and hedgehog
defects are plotted in Fig.~\ref{f-defects}a. The term which describes the
interaction of elongated bubbles is essentially given by the XY-model for which these
textures are topological excitations. Minimization of energy provides the
magnitudes of elongations $a_i$ ($a_i\sim\lambda$) and wavefunction mixing
parameters $\xi_i$, which are given in the Supplemental Information. An
important result of minimization is that textured deformation of the bubble
crystal is extended over a finite distance $L\sim 10w$ from the charge defect,
where $w\sim8\lambda$ is the lattice constant of the bubble crystal. As
distance $R_{ik}$ from the center of the charge defect increases, bubble
elongation $a_i$ decreases and the parameter $\xi_i$ in the wavefunction (2) evolves towards -1. At $R_{ik}\sim L$ the symmetric wavefunction ($\xi_i \sim 1$)
becomes energetically favorable, and round bubbles ($a_i=0$) are recovered for $R_{ik}>L$.

At very low temperatures the density of defects is low, defect separation is $>L$,
textures from different defects do not overlap, and charged defects interact
via Coulomb interaction. Consider first minimization of interaction energy for
defects of the same sign. Such defects form a superlattice superimposed on the
bubble crystal, somewhat similar to the Wigner\cite{Wigner1934} or
Abrikosov\cite{Abrikosov} lattice. Defects with opposite charges cannot easily
approach and annihilate each other due to the difference in the symmetry of
their textures, and we expect a texture crystal to form even when charge defects
of both signs are present. When temperature increases or filling factor is
shifted from the center of the RIQHE state, the density of defects increases, see
Fig.~\ref{f-defects}. When the defect separation becomes $<2L$ textures of
neighboring defects start to overlap and energy (\ref{eq:effH_pm_d}) includes
interaction between elongated bubbles belonging to different defects. This
interaction is described by the XY model. In order to calculate energy caused
by such interaction, we take the lower bound on the "exchange constant" $J$,
which is the ${\boldsymbol{\mu}}_i \cdot {\boldsymbol{\mu}}_j$ term in
(\ref{eq:effH_pm_d}). We assume that $J$ is defined by a characteristic
magnitude of elongations $a\sim\lambda$ in the region where topological
excitations overlap, which sets the lower bound to $J\sim e^2 a^2/\kappa L^3$.
The entropy of topological excitations is given then by\cite{ChaikinLubensky}
\begin{equation}
\label{BKTlog}
E=(\pi J-2T)\ln(\mathcal{L}/w),
\end{equation}
where $\mathcal{L}$ is the size of the system and the core of topological
excitations is taken to be of the order of the bubble crystal lattice constant
$w$. Such a system must exhibit the Berezinski-Kosterlitz-Thouless (BKT)
transition \cite{Berezinskii1972,Kosterlitz1973} at $T_{KT}=\pi J/2$. However,
the thermodynamic transition in the RIQHE regime differs from a conventional BKT
transition, e.g., as discussed in \cite{Minnhagen1987}, due to the finite size of
topological defects $L$. At low densities there is no overlap between
neighboring defects, and thus, the system cannot undergo the BKT transition.
The temperature $T_L$, at which textures from neighboring defects begin to overlap,
corresponding to defect density $1/(\pi L^2)$, is much higher than the estimated
$T_{KT}\approx 5$ mK. Thus, $T_{KT}$ itself is not observed experimentally. The
unbinding of topological defects at $T_L$ required to avoid the divergence of
logarithmic interactions in (\ref{BKTlog}) constitutes {\it melting of the defect
rather than the bubble crystal}. We plot $T_L$ in Fig.~\ref{f-defects}c and over
the $R_{xx}(T,B)$ data in Fig.~\ref{f-T}. The calculated $T_L$ describes the
measured temperature of the insulating-to-conducting transition remarkably
well. Following Thouless \cite{ThoulessMelting,Halperin} we estimate that
melting of an ideal bubble crystal due to dislocations takes place at
$T_m\approx 400$ mK $>T_L$. Thus we expect that topological defects continue to
exist within some temperature range above $T_L$. We conclude that the presence of topologically
non-trivial textures around charge defects is responsible for the observed
strain dependence of resistivity.

We now address the microscopic mechanism of strain response of the resistivity
tensor, and how it depends on the symmetry of topological defects. Effects of
strain $\varepsilon_{xx}=-\varepsilon_{yy}=\varepsilon$ is added to the model
similarly to the case of the stripe phase\cite{Koduvayur2011} using a deformed
coordinate system $x\rightarrow x(1+\varepsilon)$, $y\rightarrow
y(1-\varepsilon)$. In this coordinate system the lattice is identical to the
unstrained one, but the interaction potential becomes anisotropic. The linear in
strain contribution to the Hamiltonian in a real space representation can be
written as $H_{\varepsilon}=-\varepsilon \tilde{U}_f^{\epsilon}(a)
\cos[2(\phi_i-\theta_{ik})]$, where $(\phi_i-\theta_{ik})$ and $\theta_{ik}$
are polar angles of $\mathbf{a}_i$ and $\mathbf{R}_{ik}$, and $
\tilde{U}_f^{\epsilon}(a)$ is defined in the supplement. Note that $\varepsilon
a^2 \cos[2(\phi_i-\theta_{ik})]$ is precisely the $T_2$ invariant in our
symmetry argument. After minimizing the total energy we find that strain leads
to small rotations of elongated bubbles
\begin{equation}
\label{eq:phases_v_av}
\delta\phi_i=\varepsilon\gamma_i\sin[2(\phi_i-\theta_{ik})]=\pm\varepsilon\gamma_i\sin(2\theta_{ik}),
\end{equation}
where the positive sign corresponds to vortex ($\phi_i=0,\pi$) and negative to
hedgehog ($\phi_i=\pi/2,3\pi/2$) textures, $
\gamma_i=\frac{1}{\beta_i}U_f^{\epsilon}~ $. Effect of strain on defect
textures is sketched in Fig.~\ref{f-defects}.

A strain dependent contribution to conductivity arises from re-adjustment of
textures and rotation of elongated 2\={e} bubbles by an angle
$\Delta\theta_i=(\theta_{ik}-\theta_{ik'})+[\delta\phi_i(\theta_{ik})-
\delta\phi_i(\theta_{ik'})]=\Delta\theta_{i}^a+\Delta\theta_i^\varepsilon$ when
defect charge hops from site $k$ to $k'$. Overlap between the initial and the
final state of the bubbles is
$p(\Delta\theta_i)=\left<\Psi_{ik}^*|\Psi_{ik'}\right> =
1-\alpha(a_i)(\Delta\theta_i)^2$, and the conductivity
\begin{equation}
\sigma_s^{(h,v)}\propto\prod_{i}{p(\Delta\theta_i)},
\label{eq:cond}
\end{equation}
where the product includes contributions from all bubbles that rotate or change
shape during the charge hopping event. Rotation of elongated bubbles
$\Delta\theta_i^a=(w/R_{ik})\sin\left[\theta_{ik}+(1+s)\frac{\pi}{4}\right]$
leads to a strain-independent contribution to conductivity when defect shifts
by distance $w$ in the $\hat{x}$ ($s=1$) or $\hat{y}$ ($s=-1$) direction. A
strain-dependent rotation
$\Delta\theta_i^\varepsilon=\mp\gamma_i\cos(2\theta_{ik})\Delta\theta_i^a$ has
different sign for hedgehogs $(+)$ and vortices $(-)$.

Strain-induced anisotropy of resistivity tensor has different sign depending on
whether current is carried by 1\={e} or 3\={e} defects in a 2\={e} bubble
crystal. The ratio $\varrho^{(v)}/\varrho^{(h)}$ is calculated using
(\ref{eq:cond}) and is plotted in Fig.~\ref{f-strain} as a function of strain.
A corresponding ratio of resistances $R^{(v)}/R^{(h)}$ is further amplified by
the van der Pauw geometry \cite{Simon1999}. Experimentally measured strain
dependence of magnetoresistance in the vicinity of RIQHE states on the 2LL and
3LL is plotted in Fig.~\ref{f-strain}c. For each state the ratio $R^{R}/R^{L}$
is extracted, where $R^L$ and $R^R$ are resistances on the low-$\nu$ (L) and
high-$\nu$ (R) sides of the RIQHE state. Good quantitative agreement between
$R^{(v)}/R^{(h)}$ and $R^{R}/R^{L}$ for the B$_1$ state strengthen our
conclusion that 1\={e} (hedgehog) charge excitations dominate on the low-$\nu$
side of B$_1$, while 3\={e} (vortex) excitations dominate on the high-$\nu$
side. Thus our theory explains both the symmetry and unusually large magnitude of strain's effect on resistivity, in addition to
identifying the temperatures of the observed metal-insulator transitions.
Opposite strain dependence of B$_2$ underscores that the system largely
possesses an electron-hole symmetry. Surprisingly, strain dependence measured
for the I$_1$ and I$_3$ states on the 2LL is similar to the strain dependence
of the B$_1$ and B$_2$ states on the 3LL. This observation lead us to suggest
that RIQHE states observed on the 2LL originate from a crystal of defects in a
2\={e} bubble crystal interrupted by fractional quantum Hall effect states. We
note that despite the fact that Hartree-Fock approach loses its predictive
power for electron interactions on Landau levels with lowest indices, the
similarity of strain dependence of resistivity on the 2LL and 3LL suggests that
change of symmetry of topological textures with magnetic field plays an
important role in the physics of electrons even on the 2LL.

\textbf{Methods Summary}


Samples are fabricated from modulation-doped 30-nm wide AlGaAs/GaAs/AlGaAs
quantum wells grown on (001) GaAs by molecular beam epitaxy. Devices are cleaved
along (110) crystallographic direction. Eight Ohmic contacts are formed along
the edge of a sample by annealing InSn alloy at $420^\circ$C for 3 min. Devices
attached to a multilayered piezoelectric transducer with epoxy with either
$(110)$ ($yy$) or $(1\bar{1}0)$ ($xx$) aligned with the transducer polling
direction. A thin layer of metallized plastic foil is inserted between the
sample and the transducer, the foil is grounded during the experiments in order
to screen possible non-uniform electric fields. A shear strain induced in the
sample $\varepsilon=\varepsilon_{xx}-\varepsilon_{yy}\approx2.5\cdot10^{-7}
V_p$, where $V_p$ is the voltage on the transducer and
$\varepsilon_{xx}\approx-2\varepsilon_{yy}$ for the  $(xx)$ poling direction.
The actual strain $\varepsilon_{tot}=\varepsilon_{th}+\varepsilon$, where
thermally-induced strain $\varepsilon_{th}$ depends on the cooldown conditions
. The 2D electron gas is illuminated with a red LED at 10 K for 20 min. Several samples from two
different wafers with mobility $\sim 1\cdot10^7$ cm$^2$/Vs and density
$2-3\cdot10^{11}$ cm$^{-2}$ at 0.3 K were investigated, all show similar
results. Measurements are performed in a van der Paw geometry using low
frequency (17 Hz) lock-in technique with excitation current $I_{ac}=10$ nA.

\textbf{Acknowledgements}  Research was partially supported by the U.S.
Department of Energy, Office of Basic Energy Sciences, Division of Materials
Sciences and Engineering under Awards DE-SC0008630 (L.P.R.),  DE-SC0010544
(Y.L-G), DE-SC0006671 (G.S. and M.M.).

\textbf{Authors contribution} L.P.R. conceived and designed the experiments,
T.L. performed measurements. G.S. and Y.L-G  proposed explanation of the
experimental results and put forward the theory. J.W. and M. J. M. designed and grew
wafers for the experiments, and G.C. provided data from unstrained samples.
Y.L-G and L.P.R and M. J. M. contributed to writing the paper.

\clearpage

\begin{figure}
\def\ffile{f-anis}
\includegraphics[width=0.7\textwidth]{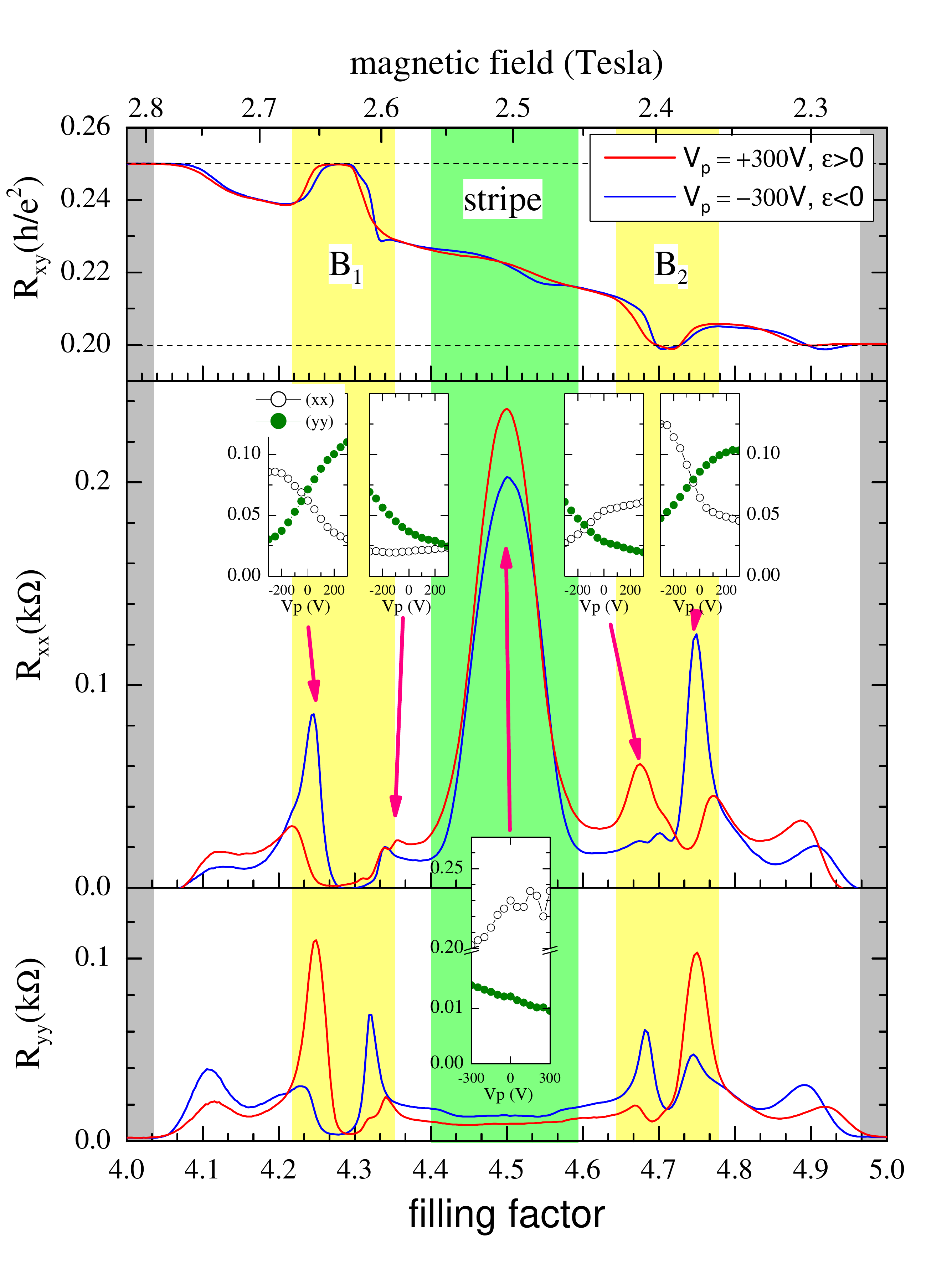}
\caption{{\bf Strain dependence of re-entrant QHE phases near filling factor
$\nu=9/2$.} Colored regions mark IQHE states (grey), RIQHE phases B$_1$, B$_2$
(yellow) and the stipe phase (green). In the inserts resistance of the peaks
marked by arrows are plotted as a function of strain for $R_{xx}$ (blue) \&
$R_{yy}$ (red). Strain dependence of the stripe phase at $\nu=9/2$ is also
shown, note the break in the vertical axis. The data is taken at 100 mK.}
\label{\ffile}
\end{figure}

\begin{figure}
\def\ffile{f-T}
\includegraphics[width=0.70\textwidth]{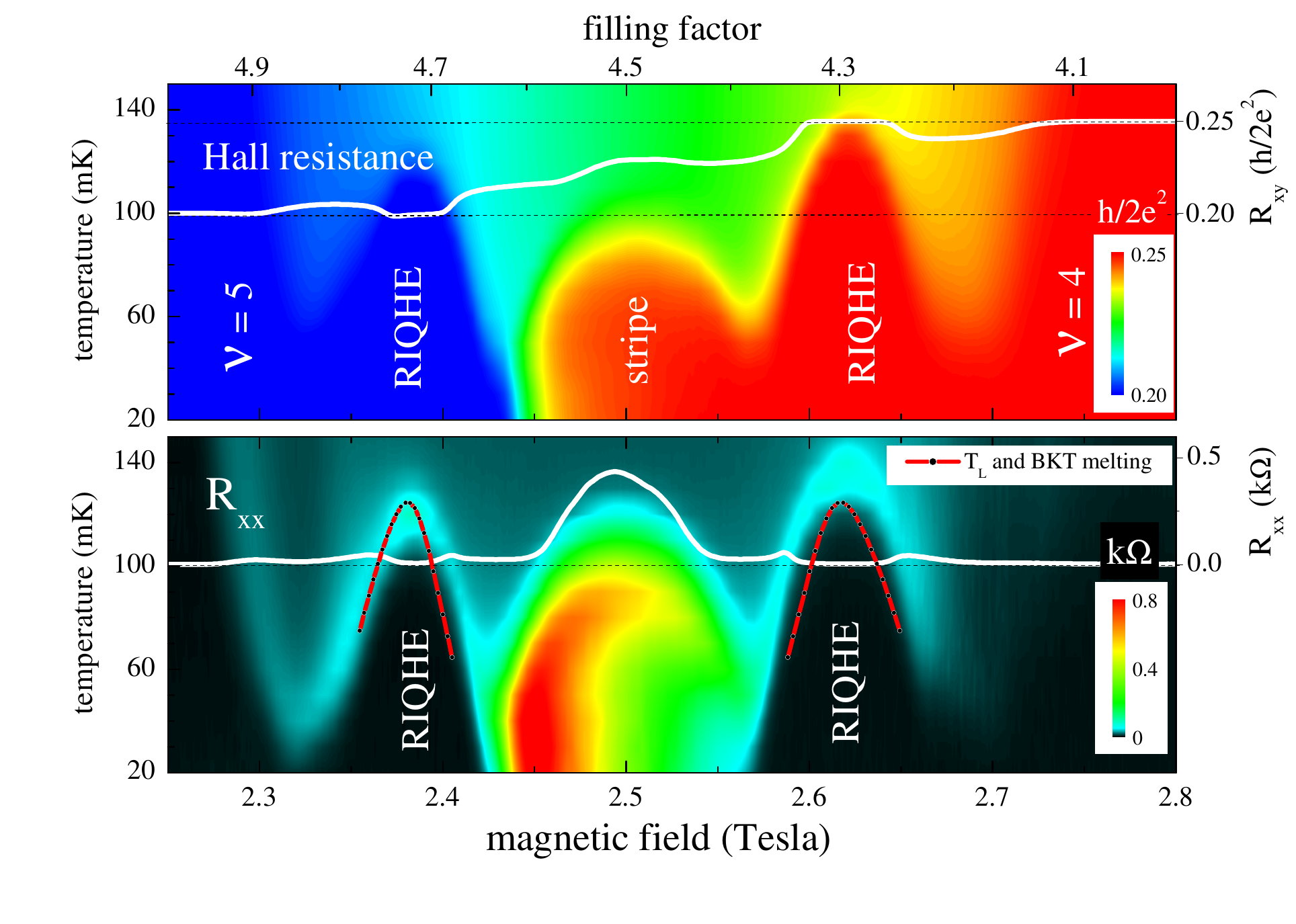}
\caption{{\bf Experimental phase diagram of melting transitions.} Calculated phase boundary
$T_L(B)$, where topological defects start to overlap resulting in melting of the defect crystal ,
is plotted on top of the experimentally measured temperature dependence of
$R_{xx}$ and coincides with the sharp increase of conductance at the boundary
of isolating and conducting phases. Color scales for $R_{xy}$ is in $h/2e^2$
and for $R_{xx}$ in k$\Omega$, the white scans are measured at $T=100$ mK.}
\label{\ffile}
\end{figure}

\begin{figure}
\def\ffile{fsym_1}
\includegraphics[width=0.49\textwidth]{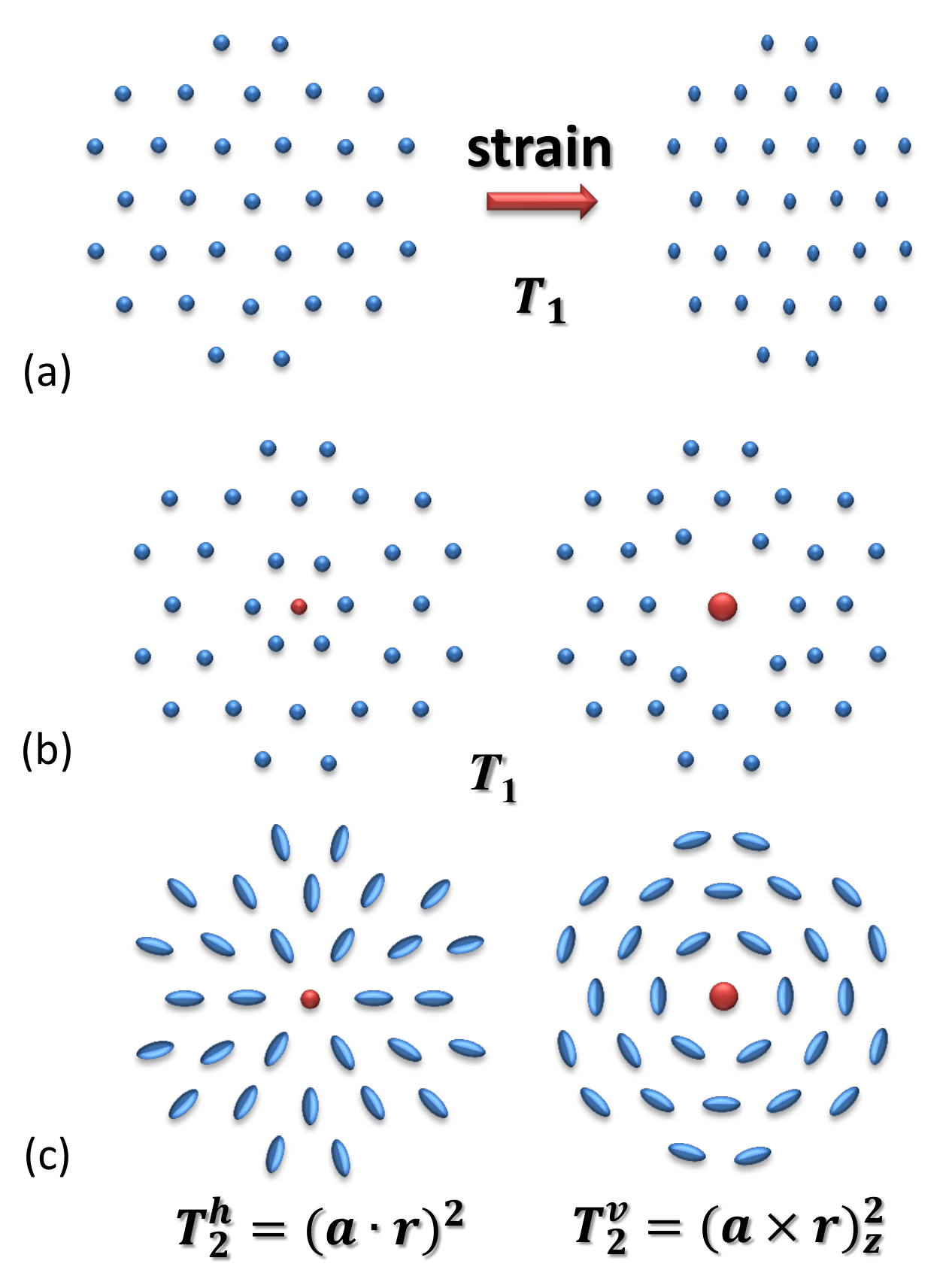}
\caption{{\bf Symmetry of bubble crystal and defects.} (a) Uniaxial strain
induces anisotropy of the effective mass and uniaxial deformation of bubble
crystal which is described by the symmetry invariant $T_1$. (b) Density of
bubbles alters in the vicinity of charged defects, but in the absence of
textures, strain response is described by the same $T_1$ invariant. (c)
functions $T^v_2=(\mathbf{a}\cdot\mathbf{r})^2$ and
$T^h_2=(\mathbf{a}\times\mathbf{r})_z^2$ give charge distributions different for vortex and hedgehog
patterns, where $\mathbf{a}$ characterizes bubble elongation and $\mathbf{r}$ -
distance between the center of the defect and elongated bubbles.}
\label{\ffile}
\end{figure}

\begin{figure}
\def\ffile{f-defects}
\includegraphics[width=0.9\textwidth]{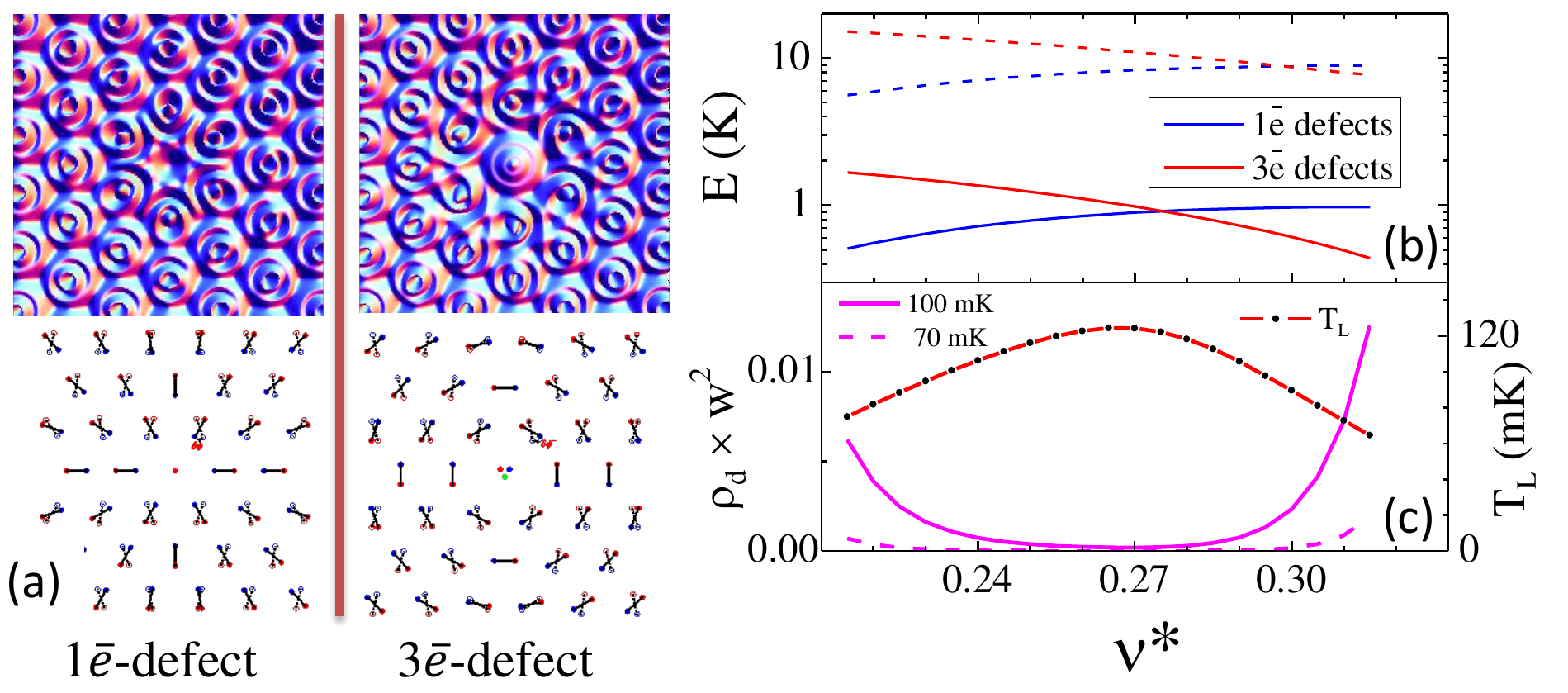}
\caption{{\bf One- and three-electron defects in a two-electron bubble
crystal.} (a) Electron density profile in the vicinity of defects. The
corresponding textures of bubble elongations are shown schematically at the
bottom. Dotted lines indicate rotations of elongated bubbles under horizontal
compressive and vertical tensile strain. (b) Activation energy for an isolated
defect calculated as a function of the filling factor of the top-most LL
$\nu^{*}$ for optimal elongations (solid lines) and for round bubbles (dashed
lines). Note that textures of elongated bubbles significantly reduce activation energies for defects. (c) Density of
defects $\rho_d$ calculated for 70 mK and 100 mK. Transition temperature
$T_L$ corresponds to the condition $\rho_d(T_L,B)=1/(\pi L^2)$. }
\label{\ffile}
\end{figure}

\begin{figure}
\def\ffile{f-strain}
\includegraphics[width=0.7\textwidth]{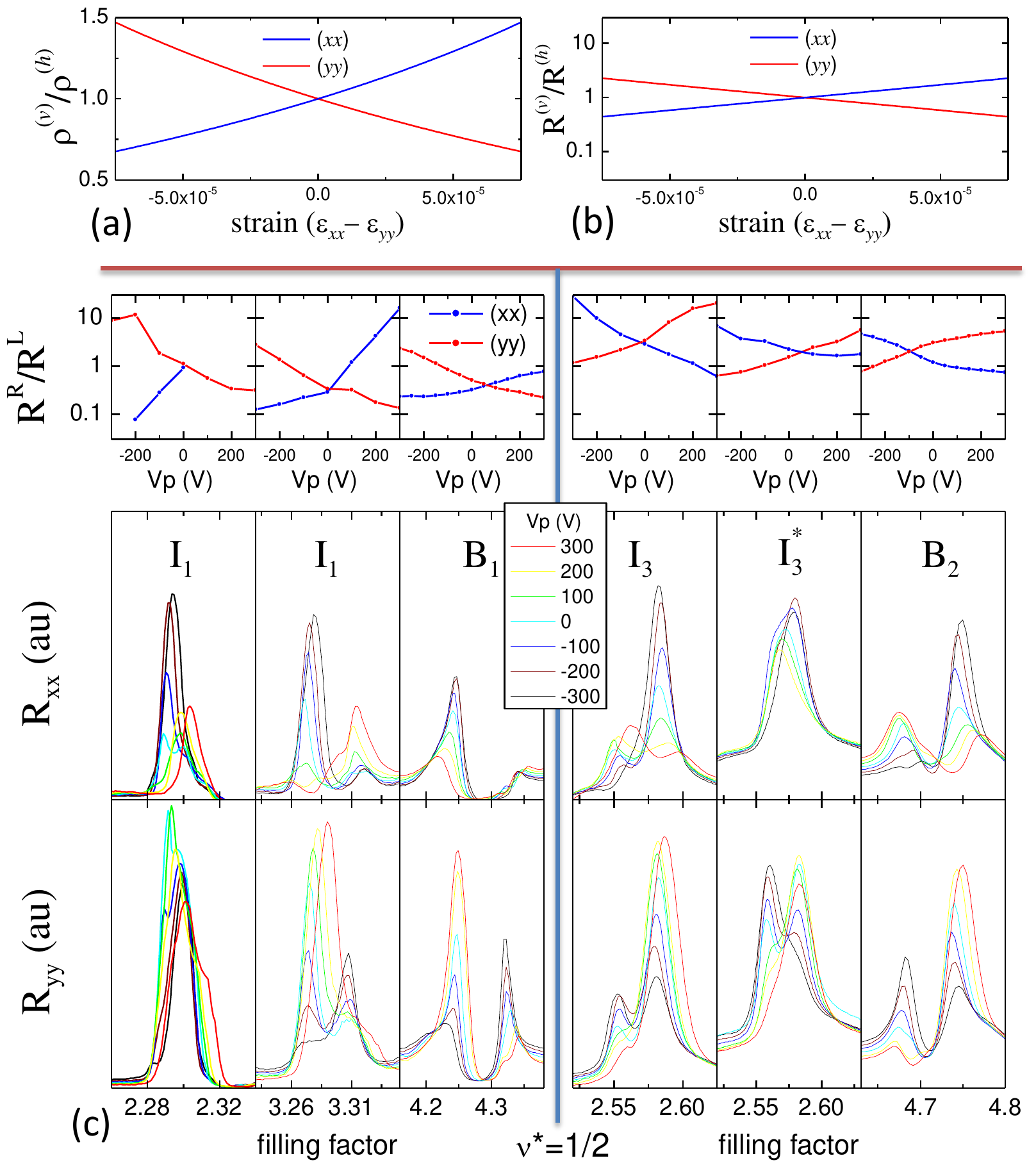}
\caption{{\bf Strain dependence of resistivity in the presence of topological
defects.} (a) Ratio of resistivities for the transport dominated by vortex
$\rho^{(v)}$ and hedgehog $\rho^{(h)}$ charge defects is calculated using
Eq.~\ref{eq:cond} and the corresponding ratio of resistances is plotted for a
square van der Pauw geometry \cite{Simon1999} in (b). (c) Magnetoresistance
near RIQHE states on 2nd \& 3rd LLs measured with different strains (voltages
on the piezo transducer $V_p$). Heights of the resistance peaks $R^L$ and $R^R$
on the left and right sides of RIQHE states are extracted from two peak
Lorentzian fits, the corresponding ratio $R_R/R_L$ is plotted at the top of
each data set.}
\label{\ffile}
\end{figure}

\clearpage

%


\clearpage
\newpage

\renewcommand{\thefigure}{S\arabic{figure}}
\renewcommand{\theequation}{S\arabic{equation}}
\renewcommand{\thepage}{sup-\arabic{page}}
\setcounter{page}{1}
\setcounter{equation}{0}
\setcounter{figure}{0}

\begin{center}
\textbf{\Large Supplementary Information} \\
\vspace{0.2in} \textsc{New topological excitations and melting transitions in quantum Hall systems}\\
{\it Tzu-ging Lin, George Simion, John D. Watson, Michael J. Manfra, Gabor A. Csathy, Yuli
Lyanda-Geller, and Leonid P. Rokhinson}
\end{center}

\section{Theory of topological excitations in the quantum Hall regime}

The purpose of the theory section of the supplement is the detailed demonstration
of how topological defects of different symmetry arise in the electron bubble
crystal, how they are influenced by external strain, and how they contribute to transport
properties of the system.

The HF Hamiltonian of partially filled $n^{\rm{th}}$ Landau level is given by
\cite{aleiner95}:
\begin{equation}
\label{eq:gen HF_hamilt_S}
H_{\rm{HF}}=\frac{1}{2}\int{\frac{d^2\bf{q}}{(2\pi)^2}
[V_{\rm{H}}({\bf{q})}-V_{\rm{F}}(\bf{q})]|\rho(\bf{q})|^2}~,
\end{equation}
where $\rho(\bf{q})$ is the projected electronic density onto the uppermost LL
and the Hartree potential and exchange potentials are, respectively:
\begin{eqnarray}
\label{eq:Hartree_pot_S}
V_{\rm{H}}({\bf{q}})&=&\frac{2 \pi}{q}e^{-q^2/2}[L_{n}(q^2/2)]^2~,\\
\label{eq:Fock_pot_S}
V_{\rm{F}}({\bf{q}})&=&2\pi\int{\frac{d^2\bf{q'}}{(2\pi)^2}V_{{\rm{H}}}({\bf{q}})
e^{-i ({\bf{q}}\times {\bf{q'}})\cdot\hat z}}~,
\end{eqnarray}
where $L_{n}$ is the $n^{\rm{th}}$ Laguerre polynomial.

A charge density wave state was proposed as the ground state for 2D systems in
in the lowest Landau level \cite{fukuyama79} even prior discovery of the quantum Hall effect. This prediction appeared not to describe the lowest Landau level, but is relevant to the two-dimensional electron liquid in higher Landau levels  where phases compete\cite{fogler-review}. Hartree-Fock (HF) studies
suggest bubble or stripe phases as possible ground
states\cite{Ettouhami2006, koulakov96,Goerbig2003,Fogler1997,Cote2003}. In the
bubble phase guiding centers of electron cyclotron orbits form a triangular
lattice. A Wigner crystal, a triangular lattice with one electron per lattice
cell, is energetically favorable at small effective filling factors
$\nu*=\nu-n_f<0.2$, where $n_f$ is the number of filled Landau levels and $\nu$
is the filling factor. For higher effective filling factors bubble phases with
more than one electron per site are the ground state. HF calculations
\cite{Ettouhami2006, Goerbig2003, Fogler1997,Cote2003} indicate the presence of
bubble phases with $M\le n+1$ electrons per bubble. The conventional picture of bubble phases is that
crystals with integer number of electrons per bubble
exist in certain range of filling factors, with lattice constant dependent on filling factor,
and first order phase transitions occur between
crystals with different number of electrons per bubble.   Density matrix
renormalization group method suggests a limit on the size of bubbles $M\le n$
\cite{Shibata2001,Yoshioka2002}. For effective filling factors close to $0.5$,
the stripe phase occurs.

Considering bubble phases, it is convenient to express the HF Hamiltonian as
the sum of effective interactions between the guiding centers:
$H_{\rm{HF}}=1/2[\sum_{i\neq j}U({\bf R}_{ij})+\sum_i U(0)]$ where $i$ and $j$
labels the nodes of the triangular lattice. The effective interaction $U$ is
given by:
\begin{equation}
\label{eq:Ueff_gen_S} U({\bf{R}}_{ij})=\int{\frac{d^2\bf{q}}{(2\pi)^2}
\rho_{i}^*({\bf {q}}) [V_{\rm{H}}({\bf{q}})-V_{\rm{F}}(\bf{q})]} \rho_{j}({\bf
{q}})\exp{(i\mathbf{q}\cdot{\bf R}_{ij})}~,
\end{equation}
where $\rho_{i}$ represents the projected density of the bubble located at site
$i$. We surmise that Hartree-Fock approach captures physics of the quantum Hall
systems even at low $n$, particularly on the 3rd LL ($n=2$) and 2nd LL ($n=1$)
\cite{Ettouhami2006,Goerbig2003,Cote2003}.

We now seek the many-body wavefunction in the range of electron densities
favorable for existence of two-electron bubbles. Experimental data leads us to consideration of one and three-electron bubble defects
in the two-electron bubble crystal. Indeed, our experiment shows a conducting state above a certain temperature. In the absence of defects,
the only possible conduction mechanism comes from melting of the bubble crystal and transition to a bubble liquid. However, in such a liquid, effect of strain on resistivity
would be similar to the effect on resistivity caused by anisotropy of the effective mass in the presence of strain, which has the same sign throughout the range of filling factors corresponding to
reentrant integer quantum Hall effect. Thus, we explore the possibility that novel defects in the two-electron bubble lattice might be favorable in the range of filling factors close the reentrant integer quantum Hall effect and are responsible for the results seen in strain experiments.

We theoretically demonstrate that it is energetically favorable for crystalline two-electron bubbles to change their shape and to become elongated. We will begin by considering how much energy it costs to create a defect in a bubble crystal, when crystalline bubbles continue to be arranged in a lattice. Next, we will assume the shape of the crystalline
bubbles to be round, but will allow their positions to re-adjust, lowering the activation energy of the defect. We then will compute the activation energy for defects taking into account elongation of crystalline bubbles.

Consider a system of 2 electron crystalline bubbles with a microscopic number
$N_d$ of infinitely separated defects that are bubbles with
$2-\sigma_d$ electrons ($\sigma_d=\pm 1$). Considering the electron system at a fixed filling factor,
we keep the total number of electrons $2N$.

The ground state is the triangular lattice of $N$ 2-electron bubble
electrons, with lattice constant $w$.  When defects are present, the
total number of bubbles has to change to $N+N_d\sigma_d/2$, as we keep the total number of electrons the same.
Assuming all bubbles to be still arranged in a triangular lattice, we find that the change in
lattice constant is $\delta w=-w \sigma_d N_d/(4N)$. Then the energy of the lattice with defects is:
\begin{equation}
\frac{1}{2}N\left[1-\left(1-\frac{\sigma_d}{2}\right) \frac{N_d}{N}
\right]\epsilon_2^{(w+\delta w)}+N_d\epsilon_d^{(w+\delta w)}~,
\end{equation}
where $\epsilon_2^{w}$ is the energy required to add one 2-electron bubble in
the bubble crystal with lattice constant $w$,  $\epsilon_d^{w}$ is the energy of a defect embedded in an
otherwise pristine system with lattice constant $w$.

From Eq. (\ref{eq:Ueff_gen_S}),  $\epsilon_2^{w}$ can be expressed as
\begin{equation}
\epsilon_2=\sum_k{\zeta_k a_k^{(2)}/w^{2k+1}}+\tilde{U}(0),
\end{equation}
where $\tilde{U}(0)=U(0)/2=833 \sqrt{\pi}/2048$ is the formation energy,
 $a_0=4$, $a_1=14$, $a_2=765/4$. Similarly,
the interaction between the defect and two-electron bubbles is
$U_d=\sum_k\sum_k{b_k^{(d)}/r^{2k+1}}$, making
$\epsilon_d=\sum_k{\zeta_k b_k^{(d)}/w^{2k+1}}+u_d$, where the formation energy of a
one-electron bubble defect is $u_{1}=0$ while for the three-electron bubble defect $u_3=77463\sqrt{\pi}/65536$.
For one electron defect we find $b_0{(1)}=2$, $b_1{(1)}=13/2$, $b_2{(1)}=117/2$, and for
three electron defect $b_0{(3)}=6$, $b_1{(3)}=45/2$ and $b_2{(3)}=531/2 $. The
parameters $\zeta$ are $\zeta_0=-4.2$, $\zeta_1=11.03$ and
$\zeta_2=6.76$.

 Using these expressions for $\epsilon_2$ and $\epsilon_d$ we find the energy of $N_d$ defects
\begin{equation}
N_d E_d=\frac{1}{2}N\left[1-\left(1-\frac{\sigma_d}{2}\right)
\frac{N_d}{N} \right] \left(\epsilon_2(w)+\frac{\partial
\epsilon_2}{\partial w}\delta
w\right)+N_d\epsilon_d(w)-\frac{1}{2}N\epsilon_2(w)
\end{equation}
The energy of one defect is:
\begin{equation}
E_d=- \frac{\sigma_d}{8} \frac{\partial \epsilon_2}{\partial w} w+
\frac{\sigma_d-2 }{4}\epsilon_2 +\epsilon_d.
\end{equation}

We now consider decrease in energy of the bubble lattice assuming the two-electron bubbles keep their round shape, but will allow positions of bubbles to deviate from triangular lattice. We approach this problem
in the spirit of ideas of Fisher,  Halperin and Morf  \cite{Halperin}. Due to the presence of a defect with charge different from charge of surrounding bubbles,
electron bubbles at lattice sites $\mathbf{R_i}$  experience displacements $\mathbf{u}(\mathbf{R_i})$ from their equilibrium  positions. In the framework of elasticity theory, the energy associated with such displacements up to the second order in $\mathbf{u}(\mathbf{R_i})$ is given by
\begin{eqnarray}
E_d(\left\{{\bf{u}}_i\right\})&=&\frac{1}{2}
 \sum_{i,j}{\Pi_{\alpha\beta}({\bf{R}}_i ,{\bf{R}}_j) u_{\alpha} ({\bf{R}_i})u_{\beta }
({\bf{R}}_j)} \nonumber\\
&-&\sum_{i\neq i_0} {\delta V^1_{\alpha} ({\bf{R}}_i)
u_{\alpha}}({\bf{R}}_i) -\sum_{i\neq i_0} {\delta V^2_{\alpha,
\beta}({\bf{R}}_i) u_{\alpha}({\bf{R}}_i) u_{\beta}({\bf{R}}_i)}
\end{eqnarray}
where $\Pi$ is the spring constant matrix,
\begin{equation}
\delta V^1_{\alpha} =\frac {\partial }{\partial r_{\alpha}}
[U_2({\bf{r}})-U_d({\bf{r}})] ~,
\end{equation}
and
\begin{equation}
\delta V^2_{\alpha, \beta} =\frac{1}{2}\frac {\partial^2 }{\partial
r_{\alpha}\partial r_{\alpha}} [U_2({\bf{r}})-U_d({\bf{r}})]~.
\end{equation}
Here the potential energy describing the bubble lattice $U_2$ is given by
\begin{equation}
\label{eq:2elebubble_int_S0}
U_2({\bf{R}}_{ij})=\frac{4}{R_{ij}}
+\frac{14}{R_{ij}^3}+\frac{765}{4R_{ij}^5},
\end{equation}
the interaction energy of the two-electron bubble at site $i$ with a one-electron defect at site k $U_{d=1}$
is given by
\begin{equation}
\label{eq:1eldef_int_S0}
U_1({\bf{R}}_{ik})=\frac{2}{R_{ik}}+ \frac{13}{2 R_{ik}^3}+ \frac{117}{2 R_{ik}^5} ~,
\end{equation}
the interaction energy of the two-electron bubble at site $i$ with a three-electron defect at site k $U_{d=3}$
is given by
\begin{equation}
\label{eq:3eldef_int_S0}
U_3({\bf{R}}_{ik})=\frac{6}{R_{ik}}+\frac{45}{2 R_{ik}^3}+\frac{531}{2 R_{ik}^5}~.
\end{equation}
 Multipole terms in expansions (\ref{eq:2elebubble_int_S0}-\ref{eq:3eldef_int_S0}) appear in magnetic field as
a result of the shape of the electron wavefunction in the second Landau level. Because the bubble crystal lattice constant is about 8 magnetic lengths, we restrict these asymptotic expansions to three terms.
 Here and throughout the rest of the supplement  we will express distances in units of magnetic length $\lambda$, wavevectors in units of
$1/\lambda$ and energies in units of $e^2/\kappa\lambda$, where $\kappa$ is the
background dielectric constant.
The spring constant matrix is determined in terms of its Fourier-transform, which is related to  Fourier-transforms of the
potentials $\delta V^1_{\alpha}$ and $\delta V^2_{\alpha,\beta}$ by
\begin{equation}
\delta V^2_{\alpha, \beta}({\bf{q}}) =
\frac{1}{2}\sum_{\gamma}V_{\gamma,\gamma}\delta _{\alpha \beta} + A_c
\Pi^d_{\alpha \beta}
\end{equation}
and
\begin{equation}
\delta V^1_{\alpha}({\bf{q}}) = -i A_c\sum_{\gamma}\frac{\partial \Pi^d_{\gamma
\gamma}}{\partial q_a}~,
\end{equation}
where $A_c$ is the elementary cell area of the bubble lattice.
Assuming a neutralizing background and writing the potential in the form
\begin{equation}
V({\bf{r}}) =\sum_{k} {\frac {a_k}{r^{2k+1}}}~,
\end{equation}
we obtain an explicit expression for $\Pi$
\begin{equation}
A_c^2\Pi_{\alpha \beta}({\bf{q}}) =2\pi\frac{q_{\alpha}
q_{\beta}}{q}+\sum_{k} {\frac {\Xi_k a_k}{w^{2k+1}} \left(
q^2\delta_{\alpha\beta} +\frac{4k+6}{2k-1}q_{\alpha}
q_{\beta}\right)}~,
\end{equation}
where $w$ is the lattice constant, and constants $\Xi_0=0.26$ , $\Xi_1=2.07$ and $\Xi_2=6.34$.
The change of energy of the lattice in terms of Fourier-transformed quantities is given by
\begin{eqnarray}
E_d(\left\{{\bf{u}}_i\right\})&=& \frac{1}{2} \int{\frac{d
{\bf{q}}}{4\pi^2} {\Pi_{\alpha \beta}({\bf{q}}) u_{\alpha}
({\bf{q}})u_{\beta } ({\bf{q}})}} \nonumber\\
&-&\int{\frac{d {\bf{q}}}{4\pi^2}
{\delta V^1_{\alpha}({\bf{q}}) u_{\alpha}}({\bf{q}})} -\int{\frac{d
{\bf{q}} d {\bf{k}} }{16 \pi^4} {\delta V^2_{\alpha,
\beta}({\bf{q}}-{\bf{k}}) u_{\alpha}({\bf{k}})
u_{\beta}(-{\bf{q}})}}.
\end{eqnarray}
 Minimization of this expression assuming
 \begin{equation}
u_{\alpha}({\bf{q}})=\frac{q_{\alpha}}{q^2}f\left(1+cq +dq^2\right)
\end{equation}
gives a decrease in activation energy of the defects caused by displacements $\mathbf{u}(\mathbf{R_i})$ .
In Fig.4b of the main paper, we plot the activation energy of one- and three-electron defects
when the displacement of round crystalline bubbles is taken into account.

Further decrease in activation energy of the defects is caused by change of shape of bubbles from round to elongated.
Until now bubbles were almost
exclusively treated as entities with uniform charge density. Ettouhami et
al.\cite{Ettouhami2004} suggested that the guiding centers in two-electron
bubbles can be spatially separated in two dimensions. We find that if a proper
superposition between wavefunctions of electrons in the same bubble is
evaluated and the phase factors due to magnetic translations are taken into
account, the energetically favorable ideal bubble crystal carries round
shaped bubbles. What we also find, however, is that in the presence of charged
defects, i.e. bubbles lacking one electron or with one extra electron, or in
the presence of charged impurities, guiding centers of electrons within bubbles
become spatially separated and two-electron bubbles become elongated. The
wavefunction of a bubble with two guiding centers separated by $2a$ can be
expressed as:
\begin{equation} \label{eq:wf_2e_bubble_gen_S}
\Psi_{\xi,\bf{a}}({\bf{r}}_1,{\bf{r}}_2)=\frac{\psi_{\bf{a}}({\bf{r}}_1,{\bf{r}}_2)
+\xi\psi_{-\bf{a}}({\bf{r}}_1,{\bf{r}}_2)}{\sqrt{2 \left(1-2e^{-2
a^2}a^2\right) \left(1+|\xi |^2\right)+4 \left(1-2 a^2\right) e^{-2 a^2} \Re
e(\xi)}}~,
\end{equation}
where
\begin{equation}
\label{eq:wf_2elebubble_S}
\Psi_{\bf{a}}({\bf{r}}_1,{\bf{r}}_2)=u_{-2}({\bf{r}}_1+{\bf{a}})
u_{-1}({\bf{r}}_2-{\bf{a}})e^{\frac{i}{2}({\bf{r}}_1-{\bf{r}}_2)\times
{\bf{a}}}-u_{-1}({\bf{r}}_1-{\bf{a}})
u_{-2}({\bf{r}}_2+{\bf{a}})e^{\frac{i}{2}({\bf{r}}_2-{\bf{r}}_1)\times
{\bf{a}}}~,
\end{equation}
where $u_{-2}$ and $u_{-1}$ are the single electron wavefunction in third Landau level ($n=2$)
in the symmetric gauge \cite{Giuliani-book}. This is a trial wavefunction similar
in spirit to the variational wavefunction proposed by Fogler and Koulakov for
round bubbles\cite{Fogler1997}.  The direction of $\mathbf{a}$ characterizes
spatial orientation of the elongated electron bubble. The electron density of
such an elongated bubble, projected on the $n=2$ LL, is
\begin{eqnarray}
\label{eq:density_2el_bubble_S} \rho_{\xi,\bf{a}}({\bf{q}})&=&
e^{-\frac{q^2}{4}} \left[\left(1-2e^{-2 a^2}a^2\right) \left(1+|\xi
|^2\right)+2 \left(1-2 a^2\right) e^{-2 a^2} \Re
e(\xi)\right]^{-1}\nonumber\\
&\times&\left\{e^{i {\bf {q}}\cdot{\bf
{a}}}\left(1+|\xi|^2-\frac{|\xi|^2q^2}{2}\right)+e^{-i {\bf {q}}\cdot{\bf
{a}}}\left(1+|\xi|^2-\frac{q^2}{2}
\right)\right.\nonumber\\
&-&e^{-2a^2+({\bf {q}}\times{\bf {a}})\cdot\hat z} \left[\left(2a^2-({\bf
{q}}\times{\bf {a}})\cdot\hat
z\right)\left(1+|\xi|^2\right)-2\left(1-2a^2\right)\Re e
\xi\right.\nonumber\\
&+&\left.\frac{\xi q^2}{2}-2\xi({\bf {q}}\times{\bf {a}})\cdot\hat
z+i\left(1-|\xi|^2\right){\bf {q}}\cdot{\bf
{a}}\right]\nonumber\\
&-&e^{-2a^2+({\bf {a}}\times{\bf {q}})\cdot\hat z} \left[\left(2a^2-({\bf
{a}}\times{\bf {q}})\cdot\hat
z\right)\left(1+|\xi|^2\right)-2\left(1-2a^2\right)\Re e
\xi\right.\nonumber\\
&+&\left.\left.\frac{\xi^* q^2}{2}-2
\xi^*({\bf {a}}\times{\bf
{q}})\cdot\hat z+i\left(1-|\xi|^2\right){\bf {q}}\cdot{\bf
{a}}\right]\right\}~,
\end{eqnarray}
where $\bar a =a_x+ia_y$, and $a^*=a_x-ia_y$.
In order to write down an asymptotic expression for the effective interaction between
bubbles at different nodes of the two-electron bubble lattice taking into account elongation of bubbles,
it is convenient
to express the interaction in terms of elongations $\mathbf{a_i}$ and effective
dipole moments $\boldsymbol{\mu}_i$, which represent rotated and re-scaled
$\mathbf{a_i}$. Retaining terms up to $R_{ij}^{-3}$ for energy change due to elongations we obtain
\begin{eqnarray}
\label{eq:2elebubble_int_S}
\delta U_2({\bf{R}}_{ij})&=&\frac{({\boldsymbol{\mu}}_i-{\boldsymbol{\mu}}_j)\cdot
 \hat{{ \bf{R}}}_{ij}} {R_{ij}^2}
+\frac{{\boldsymbol{\mu}}_i \cdot
{\boldsymbol{\mu}}_j- 3 ({\boldsymbol{\mu}}_i \cdot \hat{{
\bf{R}}}_{ij})({\boldsymbol{\mu}}_j \cdot \hat{{ \bf{R}}}_{ij})}
{R_{ij}^3}
\nonumber\\
&+&\frac{ u(a_i, \xi_i)+u(a_j,\xi_j)}{R_{ij}^3}+\frac{v(a_i, \xi_i)
\cos(2\phi_i-2\theta_{ij})+ v(a_j, \xi_j)
\cos(2\phi_j-2\theta_{ij})}{R_{ij}^3}~.
\end{eqnarray}
Here $\theta _{ij}$ is the phase of ${\bf{R}}_{ij}$, $\phi _i$ and $\phi_j$ are
the phases of elongation vectors ${\bf{a}}_i$ and ${\bf{a}}_j$ for bubbles
located at sites $i$ and $j$ respectively,
$\mathbf{\hat{R}}_{ij}=\mathbf{R_{ij}}/R_{ij}$, and
\begin{eqnarray}
\label{eq:mu_S}
\boldsymbol\mu&=&\frac{2\cal R({\bf{a}})}{e^{2
a^2}-2a^2} \frac{\sqrt{1-\left(\frac{2 \Re e \xi}
{1+|\xi|^2}\right)^2}} {1+\eta_1(a) \frac{2 \Re e \xi} {1+|\xi|^2}}~,\\
\label{eq:tilde_u_S}
u(a ,\xi)&=& a^2\frac{e^{2 a^2}+2 a^2}
{e^{2a^2}-2a^2} \frac{1+|\xi|^2-2\eta_2(a) \Re e(\xi)} {1+|\xi|^2+2\eta_1(a) \Re e(\xi)}~,\\
\label{eq:tilde_v_S}
v(a,\xi)&=&3 a^2 \frac{e^{2a^2}-2a^2+2} {e^{2a^2}-2a^2}
\frac{1+|\xi|^2+2\eta_3(a) \Re e(\xi)} {1+|\xi|^2+2\eta_1(a) \Re e(\xi)}~,
\end{eqnarray}
where the following notations have been used
\begin{eqnarray}
\label{eq:eta_eff_int_S}\eta_1(a)&= &\frac{1-2a^2}{e^{2a^2}-2a^2}~,\\
\label{eq:eta_tilde_eff_int_S}\eta_2(a)&=&\frac{1-2a^2}{e^{2a^2}+2a^2}~, \\
\label{eq:eta_bar_eff_int_S}\eta_3(a)&=
&\frac{3-2a^2}{e^{2a^2}-2a^2+2}~,
\end{eqnarray}
and $\cal R (\bf {a})$ represents vector $\bf{a}$ rotated by an angle
$\phi_{k}=\arg(1-|\xi|^2+2\Im m\xi)$.
The above analytical expressions originate in the Hartree term of the potential
(Eq.~\ref{eq:Fock_pot_S}) because the contribution to the exchange potential
(Eq.~\ref{eq:Fock_pot_S}) behaves as $\exp{(-R^2/4)}$ and is neglected.

We consider two types of charge defects, one and three-electron bubble defects, in the presence of elongations. The wavefunction of a
one-electron bubble is $u_{-2}({\bf r})$,  and has round shape. The asymptotic
expansion of the contribution to interactions between a defect at site $k$ and a two-electron
bubble at site $i$ due to elongations, with terms up to $R_{ik}^{-3}$ is obtained from Eq.~\ref{eq:Ueff_gen_S}:
\begin{equation}
\label{eq:2el1elint_S}
\delta U_1({\bf{R}}_{ik})=\frac{\boldsymbol\mu_i \hat {\bf
{R}}_{ik}}{R_{ik}^2} +\frac{ u (a_i,\xi_{i}) +
v(a_{i},\xi_i)\cos(2\phi_i-2\theta_{ik})}{2 R_{ik}^3}~.
\end{equation}

A three-electron defect, similarly to two-electron bubbles, has internal
structure and nonuniform density.  A wavefunction of a three-electron bubble is
a Slater determinant made of $u_{-2}$, $u_{-1}$, and $u_{m=0}$. Exact structure
and shape of these defects can be determined by minimizing the cohesive energy,
however, as our numerical study shows, energetics and electron transport are
primarily determined by elongations of the crystalline two-electron bubbles.
Thus, the detailed structure of the 3\={e} bubble is not essential and we model
it as having all three guiding centers on top of each other, just like
three-electron bubbles have always been treated. A contribution to effective interaction of
such a defect located at site $k$ with a two-electron bubble at site $i$ due to elongations  is given
by:
\begin{equation}
\label{eq:2el3el_int_S}
\delta U_3({\bf{R}}_{ik})=\frac{3\boldsymbol \mu_i \hat {\bf
{R}}_{ik}}{R_{ik}^2} +\frac{3\left [ u(a_i,\xi_{i}) + v
(a_{i},\xi_i)\cos(2\phi_i-2\theta_{ik})\right]}{2 R_{ik}^3}~.
\end{equation}
It is worth noting that due to the symmetry of triangular lattice $\sum_j
\cos(\phi_i-\theta_{ij})/R_{ij}^n=0$, where the summation runs over all sites of
an ideal crystal and $n$ is a positive integer. Also, the following result is
used in what follows: $\alpha=\zeta_1=w^3\sum_{j}1/R_{ij}^3\approx 11.03$, where $w$ is the lattice
constant of the bubble crystal.

If a charge defect is embedded in the otherwise pristine system, the effective
Hamiltonian  in the presence of elongations of bubbles is derived using Eqs.~\ref{eq:Ueff_gen_S} and
\ref{eq:2elebubble_int_S}-\ref{eq:2el3el_int_S} and is given by Eq.~3 of the
main paper
\begin{eqnarray}
\label{eq:effH_pm_d_S}
H^{\pm}&=&\sum_{i\neq k} \left[\mp\left[\frac{\boldsymbol \mu_i \hat
{\bf{R}}_{ik}}{R_{ik}^2}+ \frac{v_i}{2 R_{ik}^3} \cos{(2\varphi_i)}\right]+
U_f^{\pm}(i))\right]
\nonumber\\
&+&\frac{1}{2}\sum_{i \neq j;~i,j\neq k}\frac{ {\boldsymbol{\mu}}_i \cdot
{\boldsymbol{\mu}}_j- 3 ({\boldsymbol{\mu}}_i \cdot \hat{{
\bf{R}}}_{ij})({\boldsymbol{\mu}}_j \cdot \hat{{ \bf{R}}}_{ij}) }
{R_{ij}^3}~,
\end{eqnarray}
where $(+)$ corresponds to the presence of three- and $(-)$ one-electron
bubbles, $\varphi_i=\phi_i-\theta_{i,k}$, and the following notation is
introduced
\begin{eqnarray}
v_i&=&v(a_i,\xi_i),\\
U_f^{\pm}(i)&=&U_f^{\pm}(a_i,\xi_i),
\end{eqnarray}
where
 \begin{equation}
\label{eq:en_form_eff_S} U^{\pm}_{f}(a_i,\xi_i)=\left(\frac{\alpha}{w^3}
\mp\frac{1}{2R_{ik}^3}\right) u (a_i,\xi)+\frac{1}{2}U(0,a_i,\xi_i) ~,
\end{equation}
and $U(0,a_i,\xi_i)$ is defined in Eq.~\ref{eq:Ueff_gen_S}. The first two terms
of Eq.~\ref{eq:effH_pm_d_S} represent an effective one-body (one-bubble)
energy, the third one describes the formation energy at the bubble site $i$ due
to the presence of a defect,  and the fourth term represents an effective
interaction of orientations of bubbles.

In the presence of multiple defects located at sites $k$,
\begin{eqnarray}
\label{eq:effH_pm_dm_S}
H^{\pm}&=\sum_{i\neq k}& \left[\sum_k{\mp\left[\frac{\boldsymbol \mu_i \hat
{\bf {R}}_{ik}}{R_{ik}^2}+ \frac{v(a_i, \xi_i)}{2 R_{ik}^3}
\cos(2\varphi_i)\right]}+ U_f^{\pm}(a_i,\xi_i)\right]
\nonumber\\
&+&\frac{1}{2}\sum_{i \neq j;~i,j\neq k}\frac{{\boldsymbol{\mu}}_i \cdot
{\boldsymbol{\mu}}_j- 3 ({\boldsymbol{\mu}}_i \cdot \hat{{
\bf{R}}}_{ij})({\boldsymbol{\mu}}_j \cdot \hat{{ \bf{R}}}_{ij})}
{R_{ij}^3}~.
\end{eqnarray}

The ground state of a bubble system in the presence of charged defects can be
obtain by minimizing the energy (Eq.~\ref{eq:effH_pm_d_S}). Assuming that
$a/R\ll 1$, minimization of the first term provides a zeroth order result. The
ground state of a bubble lattice with one-electron defect is obtained when
$\phi_i-\theta_{ij}=0 ,\pi$, and for a three-electron defect at
$\phi_i-\theta_{ij}=\pi/2 ,3\pi/2$. These states resemble ``vortex"
(antivortex) and ``2D-hedgehog" (antihedgehog) textures. Vortex and antivortex
texture of a one-electron defect are almost degenerate at relevant
temperatures; similarly degenerate are headgehog and antihedgehog energies. It
is important to realize that the two-body interaction of
Eq.~\ref{eq:effH_pm_d_S} essentially represent an XY-model
\cite{ChaikinLubensky}. This is transparent if we take a continuum limit
$\varphi_i\rightarrow\varphi_j$, where only $\cos{(\varphi_i-\varphi_j)}$ term
is important. For the XY-model the vortex and hedgehog textures, which minimize
the $1/R^2$ interaction of the bubble system with defects, constitute
topological excitations. We note that XY-model physics characterizes dipole-dipole interactions of Eq.~\ref{eq:effH_pm_d_S} even if continuum limit is not applied \cite{PrakashHenley}.

Minimization of energy (Eq.~\ref{eq:effH_pm_d_S}) in the limit of sufficiently
large $R_{ik}$ provides the values of the separation vector $a_i$ and the
mixing $\xi_i$. The effect of the interaction term is evaluated approximating a
bubble located far away from the defect as being surrounded by bubbles with the
same parameters $a$ and $\xi$. The result of such minimization procedure is
\begin{eqnarray}
\label{eq:a_approx} a_i&=&\frac{w^2} {2^{\frac{1}{6}}
3^{\frac{1}{2}}
\alpha^{\frac{3}{2}}}\left(\frac{1}{R_{ik}}\right)^{\frac{4}{3}}-\frac{2^{7/6}w^4}{3^{5/2}
\alpha^{\frac{4}{3}}}\left(\frac{1}{R_{ik}}\right)^{\frac{8}{3}}\\
\label{eq:ci_approx}
\xi_i^{3e}&=&-1+i\frac{2 w^2}{2
\alpha^{\frac{2}{3}}} \left(\frac{1}{r}\right)^{4/3}\\
\xi_i^{1e}&=&-1+\frac{2 w^2}{2 \alpha^{\frac{2}{3}}}
\left(\frac{1}{r}\right)^{4/3},
\end{eqnarray}
where superscripts $1e$ and $3e$ refer to one- and three-electron bubbles.
In Fig.4b of the main paper, solid lines represent activation energies of one- and three-electron defects
when when decrease in energy due to elongations of crystalline bubbles is taken into account.

 An important result of minimization of the energy is that textures associated with isolated defects of the bubble
crystal are extended over a finite distance $L\sim 10w$ from the charge defect,
where $w\sim8\lambda$ is the lattice constant of the bubble crystal. As
distance $R_{ik}$ from the center of the charge defect increases, bubble
elongation $a_i$ decreases and parameter $\xi_i$ in the wavefunction (2) is
changing towards -1. At $R_{ik}\sim L$ symmetric wavefunction ($\xi_i \sim 1$)
becomes energetically favorable, and elongations $a_i=0$ (round bubbles) for
$R_{ik}>L$.

At low temperatures the density of defects is low, defect separation is $>L$,
textures from different defects do not overlap, and charged defects interact
via Coulomb interaction. Consider first minimization of the interaction energy for
defects of the same sign. Such defects form a superlattice superimposed on the
bubble crystal, somewhat similar to the Wigner\cite{Wigner1934} or
Abrikosov\cite{Abrikosov} lattice. Defects with opposite charges cannot easily
approach and annihilate each other due to the difference in the symmetry of
their textures, and we expect defect crystal to form even when charge defects
of both signs are present. When temperature increases or filling factor is
shifted from the center of the RIQHE state, the density of defects increases, see
Fig.~4c of the main text. When defect separation becomes $<2L$ textures of
neighboring defects start to overlap and energy (\ref{eq:effH_pm_d_S}) includes
interaction between elongated bubbles belonging to different defects. This
interaction is described by the XY model. In order to calculate energy caused
by such interaction, we take the lower bound on the "exchange constant" $J$,
which is the ${\boldsymbol{\mu}}_i \cdot {\boldsymbol{\mu}}_j$ term in
(\ref{eq:effH_pm_d_S}). We assume that $J$ is defined by a characteristic
magnitude of elongations $a\sim\lambda$ in the region where topological
excitations overlap, which sets the lower bound to $J\sim e^2 a^2/\kappa L^3$.
The entropy of topological excitations is given then by\cite{ChaikinLubensky}
\begin{equation}
\label{BKTlog}
E=(\pi J-2T)\ln(\mathcal{L}/w),
\end{equation}
where $\mathcal{L}$ is the size of the system and the core of topological
excitations is taken to be of the order of the bubble crystal lattice constant
$w$. Such system must exhibit the Berezinski-Kosterlitz-Thouless (BKT)
transition \cite{Berezinskii1972,Kosterlitz1973} at $T_{KT}=\pi J/2$. However,
the thermodynamic transition in the RIQHE regime differs from a conventional BKT
transitions, e.g., discussed in \cite{Minnhagen1987}, due to the finite size of
topological defects $L$. At low densities there is no overlap between
neighboring defects (their textures), and thus, the system cannot undergo the BKT transition.
Temperature $T_L$, at which textures from neighboring defects begin to overlap,
corresponds to defect density $1/(\pi L^2)$ is much higher than estimated
$T_{KT}\approx 5$ mK. Thus, $T_{KT}$ itself is not observed experimentally. The
unbinding of topological defects at $T_L$ required to avoid the divergence of
logarithmic interactions (\ref{BKTlog}) constitutes {\it melting of the defect
rather than the bubble crystal}. The $T_L$'s at different filling factors obtained in our calculations are very close to experimentally observed
temperatures of metal -insulator transitions.

By using our analytical model, we have arrived at an estimate very close to the experimental data. On one hand, this should be taken with a grain of salt.
We made a few assumptions in order to simplify our minimization procedures. In particular, we calculated first displacements of bubbles from lattice sites arising if bubbles are kept round, leading to results shown by dashed lines in Fig 4b of the main text.
 We then calculated the effect of elongations shown by solid lines in that figure assuming for simplicity that bubbles are arranged in a crystal lattice.
Furthermore, Hamiltonian (\ref{eq:effH_pm_d_S}) describing lattice bubbles around charged defects strictly speaking requires an account of additional terms with higher inverse power for bubbles very close to defects. We also discarded the effects of disorder in host heterostructure, which are known to result in lower experimentally observed activation energies as compared to theoretical predictions in quantum Hall studies. On the other hand, the analytical model we use here would provide rigorous results for bigger separations between lattice bubbles.
Furthermore, using our analytical model we can  provide
adequate analysis of the thermodynamics and transport, and in particular, correctly identify the symmetries of textures of lattice bubbles.
It would be of interest to see how these textures appear in full numerical simulation taking into account effects of disorder in heterostructure.

After melting of the defect crystal, topological defects determine the
resistivity as long as the bubble crystal is viable. The bubble crystal itself
is going to melt, e.g., due to dislocations. We estimate the melting
temperature of the bubble crystal using the Thouless formula \cite{ThoulessMelting,Halperin}:
\begin{equation}
\Gamma=\frac{4 e^2 \sqrt{\text{$\pi $n}_s}}{T \epsilon _r}=\frac{4
\sqrt{2} e^2 \sqrt{\pi }}{\sqrt[4]{3} T w
\epsilon _r},
\end{equation}
where $n_s$ is the bubble density, $w$ is the Wigner lattice constant, $e$ is
the electron charge, $\epsilon_r$ is the dielectric constant, $T$ is the melting temperature.
 The factor 4 comes from
the fact that in our case the bubbles have 2e charges. The dimensionless
parameter $\Gamma=78-130$ according to
\cite{ThoulessMelting,Hockney,Chakravarty,Morf}. Thus, we estimate the melting
temperature of the bubble crystal $T\sim 400$ mK.

{\textbf{The effect of strain}}.

 In order to evaluate the effect of strain
on the elongation of bubbles we consider bubbles located relatively far from
the defect charge, so we can neglect the term inversely proportional to
$R_{ij}^3$ in the formation energy, Eq.~\ref{eq:en_form_eff_S}. We limit
interactions to the nearest neighbors and assume that elongations $a_i$ vary
slowly with the distance from charged defects. Due to large distance between
the bubble and the defect, $\phi_i$ can be considered constant for the bubble
and its nearest neighbors. Small deviations $\delta\phi_i$ of $\phi_i$ change the total energy, which, expanding Eq.~\ref{eq:effH_pm_d_S}, we
express as $\delta E=\beta_i (\delta \phi_i)^2$, where
\begin{equation}
\beta_i=\frac{2}{R_{ik}^2}\frac{a}{e^{2 a^2}-2a^2} \frac{\sqrt{1-\left(\frac{2
\Re e\xi } {1+|\xi|^2}\right)^2}} {1+\eta_1(a) \frac{2 \Re e\xi} {1+|\xi|^2}}~.
\end{equation}

Addition of a shear strain $\epsilon_{xx}=-\epsilon_{yy}=\epsilon$ into the
model is done similarly to the case of stripe phase \cite{Koduvayur2011} using
the ``deformed coordinate system" obtained by the transformation of coordinate
system $x\rightarrow x(1+\epsilon)$, $y\rightarrow y(1-\epsilon)$. In this
coordinate system the lattice is identical to the unstrained one, but the
interaction potential becomes anisotropic:
\begin{equation}
V_{\rm{HF}}({\bf{q}})\rightarrow V_{\rm{HF}}\left(\sqrt{q_x^2(1-\epsilon)^2+q_y^2(1+\epsilon)^2}
\right) =V_{\rm{HF}}(q)+\epsilon\frac{\partial
V_{\rm{HF}}}{\partial q} q \cos(2\phi_q)
\end{equation}
The linear in strain contribution in the real space Hamiltonian can be written
as $H_{\epsilon}=-\epsilon U_f^{\epsilon}(a,\xi) \cos(2\phi_i)$ with
\begin{equation}
\label{eq:Ueff_strain}
U_f^{\epsilon}(a,\xi)=\int{\frac{d^2\bf{q}}{(2\pi)^2}
\rho_{\xi,a\hat{e}_x}^*({\bf {q}}) \frac{\partial
V_{\rm{HF}}}{\partial q} q \cos(2\phi_q)
\rho_{\xi,a\hat{e}_x}({\bf {q}})}~.
\end{equation}
Terms of the form $\sum_{j}f((R_{ij})^n)$ are unaffected by the presence of
strain due to triangular lattice symmetry. Strain induces small rotations of
the elongation of bubbles $\delta_{\phi_i}^{\epsilon}=\epsilon \gamma_i
\sin(2\phi_i)$, where $ \gamma_i=\frac{1}{\beta_i}U_f^{\epsilon}~. $

The angles describing elongation vectors become
\begin{equation}
\label{eq:pahses_v_av}\phi_i^{\pm}=\frac{3\pi}{4}
+\theta_{ik} \pm \left(\frac{\pi}{4} + \epsilon
\gamma_i\sin(2\theta_{ik})\right)~,
\end{equation}
where $(+)$ corresponds to the hedgehog and $(-)$ the vortex texture.


When the defect crystal melts due to overlap of textures,  charges are free to hop
from one site to another. However, at the same time the textured pattern has to
readjust for the new center of the defect and elongated bubbles must rotate
The superposition matrix element between the initial state of a bubble characterized by an elongation
vector ${\bf{a}}$ and the rotated one with an elongation vector $
{\bf{a}}+\delta{\bf{a}}$ is given by
\begin{eqnarray}
\label{eq:prob_rot_all}
&&p(\delta{\bf{a}})=\left<\Psi_{\bf{a}}^*({\bf{r}}_1,{\bf{r}}_2)|
\Psi_{\bf{a}+\delta{\bf{a}}}({\bf{r}}_1,{\bf{r}}_2)\right>=\nonumber\\
&&\!\left\{e^{i{\bf{a}}\times\delta{\bf{a}}-\frac{\delta
{\bf{a}}^2}{2}}\!\left[\left(1-\frac{\delta{\bf{a}}^2}{2}\right)
\left(1+|\xi|^2\right)-\Re e\xi\frac{\delta{\bf{a}}^2}{2}\right]\!
-2 e^{i\delta{\bf{a}} \times {\bf{a}}-2{\bf{b} }^2}\!\left[
{\bf{b}}^2
\left(1+|\xi|^2\right) -\Re e \xi \left(1-2   {\bf{b}}^2\right) \right]\!\right\} \nonumber\\
&&\times \left[ \left(1-2e^{-2 a^2}a^2\right) \left(1+|\xi |^2\right)+2
\left(1-2 a^2\right) e^{-2 a^2} \Re e(\xi)\right]^{-1}~,
\end{eqnarray}
${\bf{b}}={\bf{a}}+\delta{\bf{a}}/2$. For small rotations $\delta
\phi$ the rotation probability $|p|^2$ becomes
$p(\delta\phi)^2=1-p_0 (a) (\delta\phi)^2$ where
\begin{eqnarray}
\label{eq:prob_rot_small} p_0(a)&=&\left\{ 2 a^6 \left(|\xi|^2+2\Re
e \xi+1 \right)\left[\left(4 e^{2 a^2}+1\right) \left(|\xi|
^2+1\right)+4\Re e\xi\right]+2 a^2 \Re e\xi \left(|\xi|^2+4
\Re e\xi+1\right)\right.  \nonumber \\
&+& a^4 e^{2 a^2}\left[3 \left(|\xi|^2+1\right)^2-2 \Re e\xi
\left(|\xi| ^2+1\right)+4 (\Re e\xi)^2\right]+e^{2 a^2}
a^2\left[\left(|\xi| ^2+1\right)^2-2 (\Re e\xi)^2\right]\nonumber\\
&-&\left.2 a^4 \left(|\xi| ^2+3 \Re e\xi+1\right) \left(|\xi|^2+4\Re
e \xi+1\right)-e^{4 a^2}a^2 \left(|\xi|^2+1\right)
\left(2|\xi|^2+\Re e
\xi+2\right)\right\}\nonumber\\
&\times&\left[ \left(e^{2 a^2}-2a^2\right) \left(1+|\xi |^2\right)+2
\left(1-2 a^2\right)  \Re e(\xi)\right]^{-2}.
\end{eqnarray}

If a defect is displaced by a small distance $\delta$ along the $\hat x$
direction, the angle $\theta_{i,k}$ changes by $\sin\theta_{ik}/R_{i,k}\delta$,
while if the defect moves along $\hat y$ direction the change is
$-\cos\theta_{i,k}/R_{i,k}\delta$. Combining these expressions with the phase
of elongation vector given by Eq.~\ref{eq:pahses_v_av} we find the change in
angles $\phi_i$ for a hoping event
\begin{eqnarray}
\label{eq:angle_x}
\delta_x\phi_i^{\pm}&=+&[1\pm\epsilon\gamma_i\cos(2\theta_{i,k})]
\frac{w\sin\theta_{i,k}}{R_{i,k}}~,\\
\label{eq:angle_y}
\delta_y\phi_i^{\pm}&=-&[1\pm\epsilon\gamma_i\cos(2\theta_{i,k})]
\frac{w\cos\theta_{i,k}}{R_{i,k}}~.
\end{eqnarray}
Introducing these quantities into matrix elements of
Eq.~\ref{eq:prob_rot_small}, we find
\begin{equation}
\label{eq:prob_rot_small_all}
|p_s^{\pm}(\delta\phi_i)|^2=1-\frac{w^2}{R_{ik}^2}p_0 [1\pm
2\epsilon\gamma_i\cos(2\theta_{i,k})]
\sin^2\left(\theta_{i,k}+(1+s)\frac{\pi}{4}\right)~,
\end{equation}
where $s=-1$ corresponds to the hoping of the defect along $\hat x$ and $s=1$
along $\hat y$ direction.

The conductivity $\sigma_{xx(yy)}$ is proportional to the square
of the matrix element of position (displacement) operator $r_{x(y)}$, which is
proportional to the product of all $|p(\delta\phi_i)|^2$. Ratio
between peaks due to 1\={e} and 3\={e} defects
is given by
\begin{equation}
\label{eq:delta_sigma_sum} \frac{\sigma_{1e}}{\sigma_{3e}}=\prod_i
{\frac{1-\frac{w^2}{R_{i,k}^2}p_0 [1- 2\epsilon\gamma_i\cos(2\theta_{i,k})]
\sin^2\left(\theta_{i,k}+(1+s)\frac{\pi}{4}\right)}{1-\frac{w^2}{R_{i,k}^2}p_0
[1+ 2\epsilon\gamma_i\cos(2\theta_{i,k})]
\sin^2\left(\theta_{i,k}+(1+s)\frac{\pi}{4}\right)}}.
\end{equation}

We see that although strain leads to a small change of overlap between two bubbles, the effect on the motion of the whole topological defect,
dependent on overlap of many bubbles, is a giant effect. Thus, our model explains both symmetry aspects
and unusually large magnitude of the strain effect on resistivity observed experimentally.


\makeatletter
\apptocmd{\thebibliography}{\global\c@NAT@ctr 28\relax}{}{}
\makeatother

%


\clearpage
\newpage

\renewcommand{\thefigure}{E\arabic{figure}}
\renewcommand{\theequation}{E\arabic{equation}}
\renewcommand{\thepage}{e-\arabic{page}}
\setcounter{page}{1}
\setcounter{figure}{0}
\setcounter{section}{0}

\begin{center}
\textbf{\Large Extended Methods} \\
\vspace{0.2in} \textsc{New topological excitations and melting transitions in quantum Hall systems}\\
{\it Tzu-ging Lin, George Simion, John D. Watson, Michael J. Manfra, Gabor A. Csathy, Yuli
Lyanda-Geller, and Leonid P. Rokhinson}
\end{center}

\section{Thermally-induced strain.}
\label{strain}

\begin{figure}[h]
\def\ffile{f-pzt}
\includegraphics[width=0.7\textwidth]{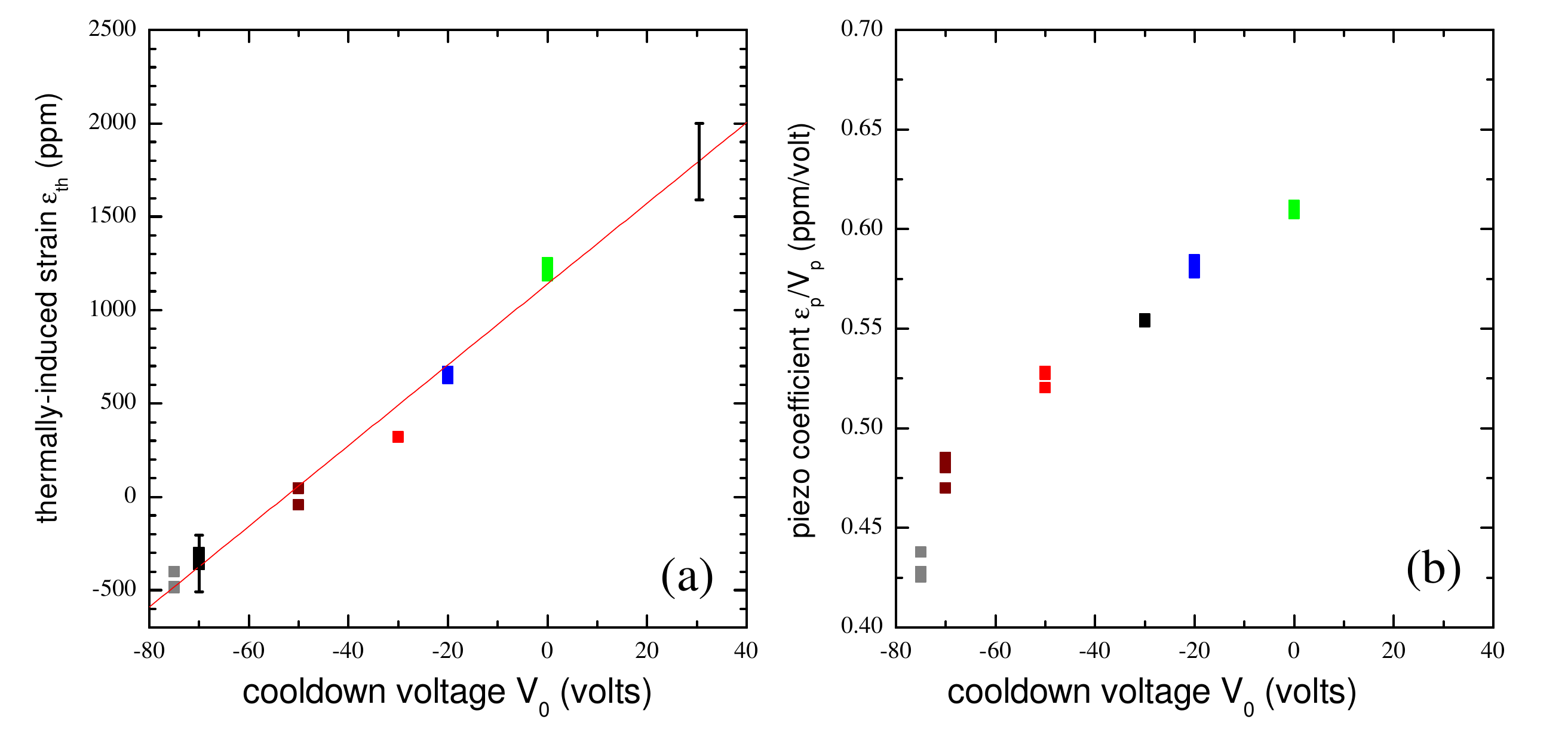}
\caption{{\bf Thermally induced strain.} Thermally induced strain can be
controlled by applying voltage $V_0$ on the transduced during cooldown. The
piezo coefficient has weak dependence on $V_0$.}
\label{\ffile}
\end{figure}

A PZT-based multilayer piezoelectric transducer has very anisotropic thermal
expansion coefficients ($+3/-1$ ppm/deg near room temperature) along and
perpendicular to the polling direction.  The residual thermally induced strain
$\varepsilon_{th}$ (at $V_p=0$) can be controlled by voltage $V_0$ applied to
the transducer during cooldown. A dependence of $\varepsilon_{th}(V_0)$ is
plotted in Fig.~\ref{f-pzt}, measured with a bi-axial strain gauge. There is
almost no thermal expension/contranction and no hysteresis in the
$\varepsilon_p(V_p)$ characteristic below 10 K.

\end{document}